\documentclass[reprint,
superscriptaddress,
amsmath,amssymb,
aps,
prb,
]{revtex4-2}

\usepackage{graphicx}% Include figure files
\usepackage{dcolumn}% Align table columns on decimal point
\usepackage{bm}% bold math
\usepackage{color}
\usepackage{hyperref}% add hypertext capabilities
\usepackage{esint}
\usepackage{xcolor}
\usepackage{soul}

\newcommand{\rr}[1]{{\textcolor{blue}{#1}}}

\newcommand{\pdag}{\phantom{dagger}}

\begin{document}

\widetext{Copyright notice: This manuscript has been authored by UT-Battelle, LLC under Contract No. DE-AC05-00OR22725 with the U.S. Department of Energy. The United States Government retains and the publisher, by accepting the article for publication, acknowledges that the United States Government retains a nonexclusive, paid-up, irrevocable, world-wide license to publish or reproduce the published form of this manuscript, or allow others to do so, for United States Government purposes. The Department of Energy will provide public access to these results of federally sponsored research in accordance with the DOE Public Access Plan (http://energy.gov/downloads/doe-public-access-plan).}

\title{Unveiling In-Gap States and Majorana Zero Modes in Superconductor–Topological Insulator Bilayer model}

\author{Umesh Kumar}
% \email{First.Author@institution.edu}
\affiliation{Materials Science and Technology Division, Oak Ridge National Laboratory, Oak Ridge, Tennessee 37831, USA}

\author{Rafa\l{} Rechci\'nski}%
\affiliation{Institute of Physics, Polish Academy of Sciences, Aleja Lotnikow 32/46, PL-02668 Warsaw, Poland}
\affiliation{Microsoft Quantum, Station Q, University of California, Santa Barbara, California 93106, USA}

\author{Tatiana de Picoli}
\affiliation{Department of Physics and Astronomy, Purdue University, West Lafayette, Indiana 47907, USA}

\author{Jukka Vayrynen}
\affiliation{Department of Physics and Astronomy, Purdue University, West Lafayette, Indiana 47907, USA}

\author{Satoshi Okamoto}
\affiliation{Materials Science and Technology Division, Oak Ridge National Laboratory, Oak Ridge, Tennessee 37831, USA}

\begin{abstract}
Interfaces between topological insulators and superconductors are promising platforms for realizing Majorana zero modes (MZMs) via the superconducting proximity effect. We introduce a bilayer model consisting of the surface states of a three-dimensional topological insulator (3DTI) coupled to an $s$-wave superconductor and systematically study the role of interlayer tunneling strength ($t_\perp$). We find that increasing $t_\perp$ shifts the proximity-induced (PrI) gap minima away from the $\Gamma$-point, giving rise to momentum-selective interference patterns that manifest as spatial oscillations in the in-gap states. By introducing an antidot with a magnetic vortex in the SC layer, we investigate the nature of in-gap states including MZMs and Caroli–de Gennes–Matricon (CdGM) modes. With increasing hybridization strength, the energy separation between MZMs and CdGM states increases, enhancing the isolation of MZMs. Importantly, in the strong hybridization limit, the leading CdGM separation remains large inspite of the decrease in the PrI gap. Spin- and spatial-resolved wavefunction analysis reveals angular momentum asymmetries absent in conventional $s$-wave systems. A direct comparison with a standalone $s$-wave superconductor confirms the emergence of distinct $p$-wave–like features in the bilayer geometry. Our results provide experimentally relevant predictions for tuning the stability of MZMs and their differentiation from the CdGM modes in SC–3DTI heterostructures and offer a theoretical framework for probing unconventional superconductivity in engineered topological systems.
\end{abstract}

\maketitle

%\tableofcontents

\section{Introduction\label{sec:Intro}}
The search for a suitable platform for topological quantum computing remains a challenge in condensed matter physics and materials science, with multiple pathways proposed with different materials and mechanisms~\cite{RevModPhys.80.1083}. The prominent major pathways are: a) superconductor (SC) and 3D topological insulator (3DTI)  surface state (SS) interfaces realizing Majorana zero modes (MZM) in materials~\cite{FuKane2008,russmann2022proximity}, 
b) spin-orbit coupled semiconductor nanowire proximitized by a SC~\cite{Lutchyn2010,Oreg2010}, 
c) quantum spin liquid in honeycomb Kitaev model~\cite{KITAEV20062,Banerjee2016, Kumar2022}, and d) topological states in moir\'e platforms~\cite{PhysRevResearch.3.L032070, Cai2023}. 
Vortices in $p$-wave superconductors have long been proposed to be a host for Majorana zero modes~\cite{MOORE1991362,Volovik1999,Ivanov2001,
Read2000,deng2020bound,ziesen2021low}, and a superconducting proximitized semiconductor surface can %the SC-3DTI interface
provide a natural pathway to realize such states~\cite{FuKane2008,Sau2010b,Alicea2010}. 
However, the materials platform for all these remains a challenging problem in the design of topological qubits, considering the inherent complexity of the materials. 
 
Following the original proposal by Kitaev to realize Majorana fermions in a $p$-wave paired  nanowire~\cite{Kitaev_2001}, one-dimensional (1D) semiconductor wires proximitized by superconductors have been intensively investigated~\cite{Lutchyn2010,Oreg2010, Lutchyn_2018}. This effort led to significant progress in the design of topological states~\cite{MSQuantum2023, Aghaee2025}. One of the major issues has been the long SC coherence length in Al, which can lead to fragile topology due to impurity scattering in Al~\cite{PhysRevB.108.085415}. %\tdp{TdP: where does the above statement comes from? Maybe we can change this phrase to "Beyond the challenge of realizing a topological phase, one has to consider the disorder in the system, as the topological gap is suppressed by the disorder in the nanowire. "}
%\ok{SO: I found a paper by Das Sarma.}

Additionally, vortex states in bulk Fe-based superconductors have been extensively explored using, for example, scanning tunneling microscopy (STM)~\cite{Kong2019, Liu2020, doi:10.1126/science.aax0274, PhysRevX.14.041039, doi:10.1126/science.aao1797}, following theoretical proposals of topological superconductivity in these materials~\cite{PhysRevLett.117.047001}. 
%Evidence for Majorana bound states was reported in the iron-based superconductor FeTe$_{0.55}$Se$_{0.45}$, using scanning tunneling microscopy (STM) to probe the vortex core under an applied magnetic field \cite{doi:10.1126/science.aao1797}.}

One of the key challenges in interpreting experimental results of STM is distinguishing the observed zero-energy states, which may also arise from conventional Andreev bound states (ABS)~\cite{PhysRevB.107.184509,PhysRevB.109.144512}, complicating the identification of genuine Majorana modes (similar to the situation in nanowires~\cite{Prada2020}).  
Furthermore, alternative interpretations in terms of the Caroli–de Gennes–Matricon (CdGM) states~\cite{doi:10.1126/science.ade0850} or the Yu–Shiba–Rusinov (YSR) states~\cite{Acta.Phys.Sin.21.75.1965,Prog.Theor.Phys.40.435.1968,Sov.Phys.JETP.29.1101.1969,PhysRevLett.126.076802} cannot be easily ruled out. %To address this ambiguity, recent studies have focused on characterization of CdGM states near zero energy in greater detail~\cite{PhysRevB.107.184509}.
%For 1D semiconductor wires, spin-polarization of the MZM has been also used to distinguish it from the trivial in-gap states~\cite{doi:10.1126/science.aan3670, Jäck2021}.  \red{JV: I would remove this last sentence since it is not related to vortices. Moreover, the references are about iron atom chains, not about 1D semiconductor wires. }

%A number of studies have proposed chiral superconductivity in materials, but the search remains controversial so far~\cite{RevModPhys.75.657,Jiao2020}. 
%Proposals to realize such phases using Floquet engineering have been reported~\cite{PhysRevLett.133.246606,Kumar2021}. 

In contrast to bulk compounds, engineered heterostructures, such as the interface of 3DTI SS and $s$-wave superconductors, have the tunability over physical parameters, and therefore more careful characterizations can be carried out. 
As one of such attempts, hybrid structures consisting of high-$T_c$ SC Fe(Te,Se) and Bi$_2$Te$_3$ were successfully synthesized~\cite{Yao2021,https://doi.org/10.1002/adma.202401809}. 
Because of the large superconducting gap  and high $T_c$, these systems open new avenues for exploring the Majorana bound states at the interface.
Recently, angle-resolved photoemission spectroscopy (ARPES) results of Fe(Te,Se)/Bi$_2$Te$_3$~\cite{Moore2023} was reported, where they show that such heterostructures contain the essential ingredients, spin momentum-locking in TI layer and the coupling between the two layers for realizing topological superconductivity. Furthermore, since the superconductivity appears even with very small Se concentrations~\cite{Moore2023}, disorder effects could be minimized~\cite{PhysRevB.105.035128}.
However, a detailed analysis has not been carried out to understand the nature of the superconducting state. 

Motivated by the recent experimental studies on the Fe(Te,Se)/Bi$_2$Te$_3$ heterostructure~\cite{Yao2021,https://doi.org/10.1002/adma.202401809,Moore2023}, 
here we investigate the proximity-induced (PrI) topological superconducting property of hybrid systems comprised of 3DTI SS and $s$-wave SC. % with a dispersing band, as is the case in real materials~\cite{doi:10.1126/sciadv.1602372}. %\rrc{Should we specify that the SC is monolayer/2D and make it an important part of the introduction?}.  
We consider a bilayer model that describes the interface between a 3DTI SS and an $s$-wave superconductor. 
The latter is defined on a two-dimensional lattice as a monolayer of Fe(Se,Te) grown on Bi$_2$Te$_3$ has been shown to superconduct~\cite{Yao2021}. Instead of adding the SC paring potential directly in the 3DTI SS, we take a more microscopic approach and include the interlayer tunneling $t_{\perp}$ which opens a proximity-induced SC gap.

We find that a large interlayer coupling $t_\perp$ shifts the PrI gap away from the $\Gamma$-point due to the finite dispersion of the superconducting layer.  This shift leads to Friedel-like oscillations in the wave functions of the states localized in the antidot as well as at the edges of the system.
%a finite Fermi momentum and induces Friedel oscillations in the chiral edge modes. \rrc{Are these actually Friedel oscillations? I think this term refers to modulations of  many-electron density near a defect.  }
%
Similar antidot systems have been studied in recent works, both in clean~\cite{ziesen2021low,deng2020bound} and disordered~\cite{PhysRevB.86.035441,ziesen2023,Rechcinski2021,Rafal2024} regimes. In these studies, the proximity-induced superconductivity is treated implicitly by introducing a momentum-independent pairing potential directly into the TI surface model. 
%\red{JV: I'm not sure if the below sentence is correct, do we have a citation for it? }
%\ok{SO: I think this is valid when the interlayer tunneling is incoherent and in-plane momentum is not conserved.}
%This approximation is valid when the TI is coupled to a bulk 3D superconductor with a Fermi momentum significantly larger than that of the TI. 
%\red{JV: We have $k_F = 0 $ in the SC so I'm not sure if the below is correct: }
%In contrast, our approach explicitly includes a 2D superconducting layer, with the Fermi momenta of the TI and SC being of comparable magnitude.  
In contrast, our theoretical setup explicitly includes a 2D superconducting layer with coherent tunneling.
In the presence of an antidot, we investigate the emergence of in-gap bound states and MZMs. Notably, the superconducting layer plays a nontrivial role in shaping the in-gap spectrum and influences the energy separation between MZMs and CdGM states. Our results demonstrate that the SC–3DTI SS interface is highly sensitive to interlayer tunneling strength, exhibiting rich and complex behavior that governs the formation, localization, and spectral characteristics of topological in-gap states.

%\red{JV: maybe need to cite \cite{PhysRevB.105.035128}}

% \rrc{We could briefly comment also on previous works concerning such antidots Refs.~\cite{ziesen2021low,ziesen2023,Rechcinski2021} and point out the differences compared to this paper. }

%\red{JV: Moved outline to the end of the section. }
The rest of the paper is organized as follows: We introduce our bilayer model in Sec.~\ref{sec:Methods}, where the real-space description incorporating an antidot and a vortex is also presented. 
%that describes the interface between a 3DTI SS and an $s$-wave superconductor,  incorporating an antidot, as detailed in Sec.~\ref{sec:Methods}. The model explicitly includes an interlayer tunneling ($t_\perp$) that governs the proximity effect. We begin 
Our discussion continues with the analysis of the PrI gap in Sec.~\ref{sec:PrIGap}, followed by a discussion of the chiral edge modes in Sec.\ref{sec:CEM}, and the in-gap bound states in Sec.\ref{sec:InGapStates}. Section~\ref{conclusions} summarizes our findings. 

\section{Methods\label{sec:Methods}}
\begin{figure}
\centering
\includegraphics[width=\linewidth]{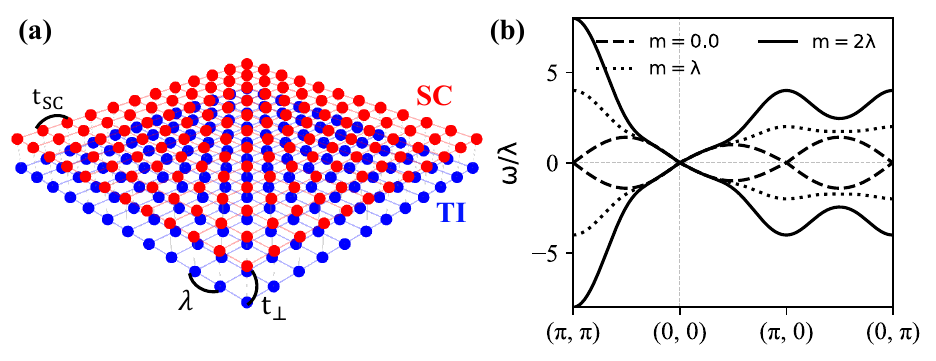}
\caption{\textbf{Bilayer square lattice of a superconductor (SC) and a topological insulator (TI) surface state.}  
(a) Schematic of the SC–TI bilayer square lattice.   (b) Bands for the 3D TI surface states for mass parameters $m=\{0, \lambda, 2\lambda\}$, highlighting the role of $m$ in opening the gap at X ($\pi$, 0), Y (0, $\pi$) and M ($\pi$, $\pi$) points in Brillouin zone.  
}
\label{fig:schematicsbilayer}
\end{figure}

We consider a bilayer system, containing surface states (SS) of a three-dimensional topological insulator (3DTI) and a conventional $s$-wave superconductor (SC), both modeled on a two-dimensional (2D) square lattice, as schematically illustrated in 
figure~\ref{fig:schematicsbilayer}~(a).  The Hamiltonian is given by $\mathcal{H} = H_\text{TI} + H_\text{SC} + H_\text{hyb}$.

The first term $H_\text{TI}$ describes the 3DTI SS given by~\cite{PhysRevX.7.031006} %\rrc{Doesn't \cite{FuKane2008} use a continuous model?}
%\ok{\bf please check my edition to this equation}
\begin{equation} \label{eq:TIHamiltonian}
\begin{split}
H_\text{TI} &=
i \frac{\lambda}{2}\sum_{{\bf r}, \boldsymbol{\alpha}} \psi_{\bf r}^\dagger \boldsymbol{\sigma} \cdot (\hat{\bf z}\times \boldsymbol{\alpha})\psi_{{\bf r} +\boldsymbol{\alpha}} \\
%i \frac{\lambda}{2}\sum_{{\bf r}, \alpha}  \psi_{\bf r}^\dagger \sigma_{\bar\alpha} \psi_{{\bf r} +\alpha} - \text{h.c.} 
&+\sum_{\bf r}  \psi_{\bf r}^\dagger \Big(\frac{3}{2} m\sigma_z + \varepsilon_\text{TI} - \mu \Big) \psi_{\bf r}^{\pdag} \\ 
& -\frac{m}{8} \sum_{{\bf r},\boldsymbol{\alpha}} \psi_{\bf r}^\dagger \big( 4 \sigma_z \psi_{{\bf r}+\boldsymbol{\alpha}}-\sigma_z \psi_{{\bf r}+2 \boldsymbol{\alpha}} \big).
\end{split}
\end{equation}
Here, $\psi_{\bf r} = \begin{bmatrix} c_{{\bf r}\uparrow} & c_{{\bf r}\downarrow} \end{bmatrix}^T$ is the spinor at the lattice site ${\bf r} = (x, y)$ in the 3DTI and $\boldsymbol{\sigma}=(\sigma_x,\sigma_y,\sigma_z)$ is the vector of Pauli matrices.  $\boldsymbol{\alpha} =  \pm \hat{\bf x}, \pm \hat{\bf y}$ denotes the vector of nearest neighbors in units of the lattice constant;  $\hat{\bf x}$, $\hat{\bf y}$, and $\hat{\bf z}$ being the unit vectors along the $x$, $y$, and $z$ directions, respectively. The parameter $\lambda$ represents the Rashba spin-orbit coupling (also equal to the Dirac velocity). %\red{JV: not sure if we should introduce the following momentum space description (it is confusing since (1) is in real space): }which preserves the form $\lambda {\bm \sigma} \cdot (\hat{\bf z}\times {\bf k}) ~=\lambda(\sigma_y k_x - \sigma_x k_y)  $. 
$\varepsilon_\text{TI}$ is the onsite potential for the TI layer, and $\mu$ is the chemical potential. 
%\red{Add more explanation about this (we circumvent Nielsen-Ninomiya, break TRS, the approximation is valid near the dirac point etc.):}

The parameter $m$ introduces a mass term that lifts the degeneracy in the surface-state spectrum of the topological insulator (TI), yielding a single gapless Dirac cone at the $\Gamma$-point, as shown in the band structure shown in figure~\ref{fig:schematicsbilayer}~(b) for a set of $m$ values. 
%The corresponding band structure reveals that the $m$ term opens gaps at all Dirac points in the Brillouin zone, except at $\Gamma$. 
This mass term is essential for isolating a single Dirac cone and thereby capturing the low-energy physics of a 3D TI surface. Notably, the inclusion of $m$ circumvents the fermion doubling problem, which states that a lattice model preserving time-reversal symmetry (TRS) cannot host an odd number of Dirac points, as dictated by the Nielsen–Ninomiya theorem~\cite{NIELSEN1981219}.

%Here, $\psi_{\bf r} = \begin{bmatrix} c_{{\bf r}\uparrow} & c_{{\bf r}\downarrow} \end{bmatrix}^T$ is the spinor at the lattice site
%${\bf r} = (x, y)$. \tdp{The Hamiltonian has the Rashba spin-orbit coupling $\lambda$, Pauli matrices $\boldsymbol{\sigma}=(\sigma_x,\sigma_y,\sigma_z)$ acting on the spin space, and $\boldsymbol{\alpha} =  \pm \hat{\bf x}, \pm \hat{\bf y}$, with $\hat{\bf x}$, $\hat{\bf y}$, and $\hat{\bf z}$ being the unit vector along the direction $x$, $y$, and $z$, respectively. The 
%TI Hamiltonian has a mass term $m$ that} lifts the degeneracy in the TI spectrum, producing a single Dirac cone at the $\Gamma$-point.

The second term $H_\text{SC}$ describes the SC layer, %. We consider $s$-wave pairing given by
\begin{equation}\label{eq:HamSC}
\begin{split}
H_\text{SC} &= t_\text{SC} \sum_{\bf r, \alpha} \xi_{{\bf r}+\alpha}^\dagger \xi_{\bf r} + (\varepsilon_\text{SC} -\mu) \sum_{\bf r} \xi_{\bf r}^\dagger \xi_{\bf r} \\ %+ \text{h.c.} 
&+ \sum_{\bf r} \bigl( \Delta_{\bf r} \xi_{\bf r}^\dagger  \sigma_y  \xi_{\bf r}^\dagger + \text{h.c.}\bigr),
\end{split}
\end{equation}
where $\xi_{\bf r} = \begin{bmatrix} d_{{\bf r}\uparrow} & d_{{\bf r}\downarrow}\end{bmatrix}^T $ and the $\Delta_{\bf r}$ is the $s$-wave SC pairing potential. The parameters $\varepsilon_\text{SC}$ and $t_\text{SC}$ are the onsite potential and hopping amplitude (related to effective mass) in the SC lattice. 

The last term $H_\text{hyb}$ describes the coupling between the two layers as a direct hopping of electrons given by
\begin{equation}
\begin{split}
 H_\text{hyb} & = t_\perp \sum_{\bf r} \bigl( \xi_{{\bf r}}^\dagger \sigma_0 \psi_{{\bf r}}^{\phantom\dagger} + \text{h.c.} \bigr),
\end{split}
\end{equation}
where $t_\perp$ denotes the interlayer tunneling amplitude, and $\sigma_0$ is the $2 \times 2$ identity matrix, thus the hybridization is diagonal in the spin index. %\red{JV: label $\sigma$ should be removed above if you use matrix $\sigma_0$.}\\

We use the Bogoliubov-de-Gennes (BdG) formalism~\cite{Zhu2016-jo, PhysRevB.103.064508}, in which each lattice site will be described by an 8-dimensional Hamiltonian (2 TI particle + 2 TI hole + 2 SC particle + 2 SC hole).  Therefore, the full Hamiltonian is given by 
$\mathcal{H} = \sum_{\bf r, \mathbf{\alpha}} \phi_{\bf r}^\dag \,  h_{\bf r, r+\mathbf{\alpha}} \, \phi_{{\bf r} +\mathbf{\alpha}}$, where $\phi_{\bf r} =  \begin{bmatrix} \psi_{\bf r} & \psi_{{\bf r}}^\dagger & \xi_{\bf r} & \xi_{\bf r}^\dagger  \end{bmatrix}^T$ in the Nambu representation. Here $\mathbf{\alpha} = 0$ accounts for the on-site and SC pairing term and the interlayer hybridization, whereas $\mathbf{\alpha} = \pm\hat{\bf x},\pm \hat{\bf y}$ for the nearest neighbor hopping terms, and $\mathbf{\alpha} = \pm\hat{2 \bf x}, \pm \hat{2\bf y}$ for next nearest neighbor hopping terms.
%\red{JV: replace $\alpha$ above by $\delta$ to be consistent? }
\\

\begin{figure}
\centering
\includegraphics[width=\linewidth]{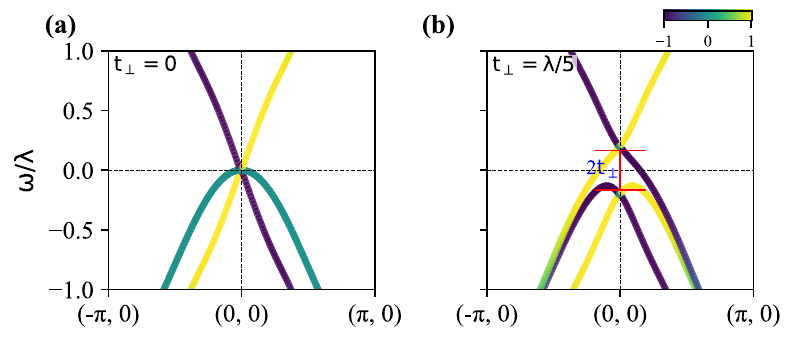}
\caption{\textbf{Hybridization of the SC-layer metallic band with the 3DTI surface state.}  
(a) Band structure of the unhybridized system showing the 3DTI surface state and the metallic band from the superconducting (SC) layer.  
(b) Hybridized band structure for interlayer coupling $t_\perp = \lambda/5$.  
The color scale indicates spin polarization along the $y$-direction 
$S_y^l(\bf k)$, as defined in  Eq.~(\ref{Eq:Sreslvd}).
}
    \label{fig:Metal3DTI_int}
\end{figure}

Throughout this work, we use the following parameters unless otherwise stated: lattice spacing as the unit length, $\lambda = 1$ as the unit energy and consider $\mu=0$, $\varepsilon_\text{TI} =0$, and $m = 2\lambda$ for the TI layer; $\Delta_{\bf r}=\Delta_0 = \lambda/5$, $t_\text{SC} = 2\Delta_0$, and $\varepsilon_\text{SC} =- 4 t_\text{SC}$  for the SC layer.   %\uk{We set the lattice constant (a) as a unit length. } \red{JV: If $a$ does not appear anywhere then I would erase "(a)" from the previous sentence. }
%\red{JV: We also consider lattice constant equal to one. }
The parameter values are chosen to achieve a qualitative agreement with the electronic structure of Fe(Te,Se)/Bi$_2$Ti$_3$ interface~\cite{Moore2023}, while also simplifying the model to facilitate efficient numerical calculations. In particular, by fixing $t_\text{SC}>0$, we ensure that when the SC layer is in non-superconducting state, the low-energy band near the Fermi level has a negative effective mass [see figure~\ref{fig:Metal3DTI_int}~(a)], consistent with the hole-like band observed near $\Gamma$ in Fe(Te,Se)~\cite{doi:10.1126/sciadv.1602372}. 
$\varepsilon_\text{SC}$ is chosen such that the band top at the $\Gamma$ point in the SC layer touches the Fermi level, which is globally set to zero energy.

To understand the interplay between the real and momentum space and present a comprehensive picture, we transform the bilayer Hamiltonian to the momentum space. The Hamiltonian is now given by $\mathcal{H} = \sum_{\bf k} \phi_{\bf k}^\dag \,   \hat h_{k} \, \phi_{\bf k} $, where $\phi_{\bf k} =  \begin{bmatrix} \psi_{\bf k} & \psi_{-{\bf k}}^\dag & \xi_{\bf k} & \xi_{-{\bf k}}^\dag  \end{bmatrix}^T$ and $\hat h_{k} = $ 
%
%\onecolumngrid
\begin{equation}\label{eq:BdGkspace}
\begin{bmatrix}
 {\bm d}_\text{TI} ({\bf k}) \cdot {\bm \sigma} & 0  & \sigma_0 t_\perp & 0\\
0 &  -{\bm d}_\text{TI} ({-\bf k}) \cdot {\bm \sigma}^*   & 0 &  - \sigma_0 t_\perp \\
\sigma_0 t_\perp & 0 & {\bm d}_\text{SC} ({\bf k}) \cdot {\bm \sigma}&  i\sigma_y \Delta \\
0 & -\sigma_0 t_\perp & -i \sigma_y \Delta^{*} &  -{\bm d}_\text{SC} ({-\bf k}) \cdot {\bm \sigma}^*
\end{bmatrix}.
\end{equation}
Here, %\rrc{I've added $-\mu$. }
$ {\bm d}_\text{TI} ({\bf k}) = (\varepsilon_\text{TI}-\mu, -\lambda \sin k_y , \lambda \sin k_x, M_{\bf k})$, ${\bm \sigma} = (\sigma_0, \sigma_x, \sigma_y, \sigma_z)$ with ${\bf k} =(k_x, k_y)$, $M_{\bf k} = m [ (2-  \cos k_x -  \cos k_y ) - \tfrac{1}{4} ( 2- \cos 2k_x- \cos 2k_y ) ] $, and $\Delta =\Delta_0$.  

In the limit ${\bf k} \rightarrow \Gamma$, $ {\bm d}_\text{TI} ({\bf k}) \cdot {\bm \sigma} = 
\begin{bmatrix}
\varepsilon_\text{TI}-\mu & -i\lambda k_- \\
i\lambda k_{+} & \varepsilon_\text{TI}-\mu
\end{bmatrix}$, where $k_{\pm} = k_x \pm i k_y $ and $|{\bf k}|^2 = k_x^2+k_y^2$. Therefore, the model preserves the TRS in this momentum region as $M_{\bf k}$ vanishes up to the cubic term. The  energies are given by  $\omega_\text{TI}^{\pm} ({\bf k}) = \varepsilon_\text{TI}-\mu \pm \lambda|{\bf k}|$, in the vicinity of the $\Gamma$ point. The eigenstates are given by $|V_\text{TI}\rangle_\pm  = \big(\sqrt{k_{-}} c_{{\bf k}\uparrow}^\dagger \pm i \sqrt{k_+} c_{{\bf k}\downarrow}^\dagger \big) |0\rangle$.
%$|V_{TI}\rangle_\pm = \begin{bmatrix} \pm \sqrt{k_-} c_{{\bf k}\uparrow} & \sqrt{k_+} c_{{\bf k}\downarrow} \end{bmatrix}^T$ \rrc{Should the ket include the annihilation operators or just the amplitudes? Just the amplitudes seem more consistent with~\eqref{Eq:Sreslvd}  }. 
For ${\bf k}$ along $(0,0)$ to $(\pi,0)$, the  $\omega_\text{TI}^{+}~(\omega_\text{TI}^{-})$ dispersion have $\uparrow$~($\downarrow$) spin along the $y$-direction, where $\sigma_x$ and $\sigma_z$ vanish as can be seen in figure~\ref{fig:Metal3DTI_int}(a).

The SC part is given by, ${\bm d}_\text{SC} ({\bf k}) = [\varepsilon_\text{SC}-\mu+2 t_\text{SC} \cos(k_x)+ 2 t_\text{SC} \cos(k_y),0,0, 0]$.  
In the limit ${\bf k} \rightarrow \Gamma$, ${\bm d}_\text{SC} ({\bf k}) \cdot {\bm \sigma} = 
\begin{bmatrix}
\varepsilon_\text{SC}-\mu + 4t_\text{SC} - t_\text{SC} |{\bf k}|^2 & 0 \\
0 & \varepsilon_\text{SC}-\mu + 4t_\text{SC}- t_\text{SC} |{\bf k}|^2
%\tilde{\mu}_d - \frac{1}{2} t_d |{\bf k}|^2 & 0 \\
%0 &\tilde{\mu}_d - \frac{1}{2} t_d |{\bf k}|^2
\end{bmatrix}$. %, where $\tilde{\mu}_d =\mu_d + 4t_d$. 
Using the condition $\varepsilon_\text{SC}-\mu = -4t_\text{SC}$ considered in our model, the SC bands are given by $\omega_\text{SC}^{\pm} ({\bf k}) = \pm \sqrt{|\Delta|^2 + t_\text{SC}^2 |{\bf k}|^2}$ and will be gapped by the pair-potential $\Delta$.%open a gap. %As can be see Fig.~\ref{fig:bands}~(d)), the gap is therefore given by $\Delta_0$ in this momentum region. 

The spin polarization in each band is evaluated using the relation
\begin{equation}\label{Eq:Sreslvd}
S_\alpha^{l}({\bf k}) = \langle \phi_{\bf k} | \sigma_\alpha^l | \phi_{\bf k}\rangle.
\end{equation}
Here, $l$ corresponds to each particle and hole index of the 3D TI and SC layer and $\alpha \in \{x, y, z\}$ components of the spin.
%\red{JV: Does $S_\alpha^{l}({\bf k})$ actually appear anywhere in the text? Not in Figs. 2-3 at least... If it doesn't, no need to introduce this notation. In the TI paragraph above you can write $\langle V_{TI} | \sigma_\alpha | V_{TI}\rangle$  to give a bit more detail.}
\\

\subsection{Model with a vortex\label{sec:Methods_vortex}}

The main focus of this work is to  study the inhomogeneous situation where the SC layer has an antidot structure [see figure~\ref{fig:Vortexmode} (a)] pierced by a magnetic flux quantum creating a vortex configuration in the SC order parameter. 
We consider the  vector potential given by 
\begin{equation}\label{Eq:VecPot}
\textbf{A}({\bf r}) = 
\begin{cases}
\begin{aligned}
&\,\frac{\Phi}{2\pi R^2}(-y, x, 0) & \text{for}~r < R, \\
&\,\frac{\Phi}{2\pi r^2}(-y, x, 0) & \text{for}~r \geq R,
\end{aligned}
\end{cases}
\end{equation}
describing a  magnetic flux  $\Phi =N \Phi_0~(= N \frac{h}{2|e|}$, where $-|e|$ is the electron charge) permeating the system, with the magnetic field ${\bf B} = (0, 0, B_0)$ distributed uniformly inside a disk of radius $R$, coinciding with the antidot radius~\cite{Rafal2024}. We assume here that the penetration depth is negligible compared to $R$. Throughout our work, we restrict to the cases of $N =0$ and 1.

%The finite $\textbf{A}({\bf r})$ outside the vortex leads to the Aharanov-Bohm effect~\cite{PhysRev.115.485}. \rrc{It does, but is it relevant?} \rr{\st{In our model with vortex, we consider vortex traps a single magnetic flux quantum given by $\Phi_0 = \frac{\hbar}{2|e|}$, where $-|e|$ is the electron charge.}}

In the above tight-binding model, the vector potential leads to the following modification. The hopping terms attain a phase accounted for by the vector potential as a Peierls substitution given by 
\begin{equation}
c_{\bf r}^\dagger c_{\bf r'}^{\phantom\dagger} \rightarrow c_{\bf r}^\dagger c_{\bf r'}^{\phantom\dagger} \mathrm{e}^{-i\frac{|e|}{\hbar} \int_{\bf r}^{\bf r'} {\bf A}({\bf s}) 
\cdot d{\bf s}} = c_{\bf r}^\dagger c_{\bf r'}^{\phantom\dagger} \mathrm{e}^{i N \Phi_{\bf r \bf r'}}
\label{eq:Peierls}
\end{equation}
in both layers due to the magnetic field.  
The  phase has analytical expressions given by  %is given by
%$ c_{{\bf r}_i}^\dagger c_{{{\bf r}_i+\delta{\bf r}}}^{\phantom\dagger} \rightarrow c_{{\bf r}_i}^\dagger c_{{{\bf r}_i+\delta{\bf r}}}^{\phantom\dagger} e^{-%i\frac{e}{\hbar} \int_{\delta {\bf r}} {\bf A}({\bf r}) 
%\cdot d{\bf r}}$ in both layers due to the magnetic field.  The phase is given by
%
\begin{equation}
\Phi_{\bf r \bf r'} =
\begin{cases} 
 \begin{aligned}
& - (x' -x) y \frac{1}{2 R^2} & \text{for}~ r,r' < R \\ 
& - \frac{1}{2} \bigg\{ \tan^{-1}\biggl(\frac{x'}{y}\biggr)-\tan^{-1}\biggl(\frac{x}{y}\biggr) \bigg\} & \text{for}~ r,r' \ge R \\
 \end{aligned}
\end{cases} \\ 
\end{equation}
for $\bf r$ and $\bf r'$ along the $x$ direction, and 
\begin{equation}
\Phi_{\bf r \bf r'} =
\begin{cases} 
 \begin{aligned}
&  +(y' -y) x \frac{1}{2 R^2} & \text{for}~ r,r' < R \\ 
& + \frac{1}{2} \bigg\{ \tan^{-1}\biggl(\frac{y'}{x}\biggr)-\tan^{-1}\biggl(\frac{y}{x}\biggr) \bigg\} &  \text{for}~ r,r' \ge R 
 \end{aligned}
\end{cases} 
\end{equation}
for $\bf r$ and $\bf r'$ along the $y$ direction, and ${\bf r} = (x,y)$ and  ${\bf r}' = (x',y')$. %\uk{The total flux through the vortex is given by $\Phi = \oint_{r<R} \Phi_{\bf rr'} \, d {\bf r} $.}

The SC layer is absent in the antidot [see figure~\ref{fig:Vortexmode} (a)], whereas Peierls phase outside the vortex has the same form as the TI layer. Additionally, the SC pairing in this layer is given by, %\rrc{Added $N$ below, for the same reason as above.}
\begin{equation} \label{Eq:SCGap}
\Delta_{\bf r} =
\begin{cases} 
 \begin{aligned}
&  0 \qquad \qquad  r < R \\ 
& \Delta_0 \mathrm{e}^{i N \varphi} ~\quad r \ge R .
 \end{aligned}
\end{cases} 
\end{equation}
Here $(r, \varphi)$ are the polar coordinates of the vortex, and $\Delta_0$ is SC gap of the superconductor. Our choice of the phase factor is detailed in Appendix~\ref{sec:PhaseFactors}. 
\\

%\red{JV: Do we need this subtitle below?} {\sl Probability Density:---} 
We characterize the chiral and bound Majorana fermions using probability density. The $m^\text{th}$-eigenvector of the real-space Hamiltonian is given by %$| \Psi_m \rangle  = \sum_{\bf r} |\Psi_m ({\bf r}) \rangle $ with $|\Psi_m ({\bf r}) \rangle = \Big(\begin{bmatrix}   u_{{\bf r}\uparrow}^\text{TI} &  u_{{\bf r}\downarrow}^\text{TI} & v_{{\bf r}\uparrow }^\text{TI}  & v_{1{\bf r}\downarrow}^\text{TI} \end{bmatrix} \circ \psi_m (r) \mathbin\Vert \begin{bmatrix} u_{{\bf r}\uparrow}^\text{SC} &  u_{{\bf r}\downarrow}^\text{SC} & v_{{\bf r}\uparrow }^\text{SC}  & v_{1{\bf r}\downarrow}^\text{SC} \end{bmatrix} \circ \xi_m({\bf r}) \Big)^\dag  | 0 \rangle$. 
$|\Psi_m ({\bf r}) \rangle = \left( u_{{\bf r}\uparrow}^\text{TI} c_{\mathbf{r}\uparrow}^\dagger+
 u_{{\bf r}\downarrow}^\text{TI} c_{\mathbf{r}\downarrow}^\dagger +
 v_{{\bf r}\uparrow}^\text{TI} c_{\mathbf{r}\uparrow} + v_{{\bf r}\downarrow}^\text{TI} c_{\mathbf{r}\downarrow} \\ +  \ldots + v_{{\bf r}\downarrow}^\text{SC} d_{\mathbf{r}\downarrow}
 \right) |0\rangle$.   Here $u$ and $v$ are the particle and hole components of the corresponding eigenvector of the Hamiltonian.

Using this eigenvector, the site-resolved probability density for a state ($m$) on TI is given by $ n_\text{TI}^m ({\bf r}) =  |u_{{\bf r}\uparrow}^\text{TI}|^2 +  |u_{{\bf r}\downarrow}^\text{TI}|^2
+ |v_{{\bf r}\uparrow }^\text{TI}|^2 + |v_{{\bf r}\downarrow }^\text{TI}|^2 $. Similarly, the probability density on SC can be evaluated as  $ n_\text{SC}^m ({\bf r}) = |u_{{\bf r}\uparrow}^\text{SC}|^2 +  |u_{{\bf r}\downarrow}^\text{SC}|^2
+ |v_{{\bf r}\uparrow }^\text{SC}|^2 + |v_{{\bf r}\downarrow }^\text{SC}|^2 $.

\section{Results and Discussion\label{sec:ResultsDiscussion}}

We begin by analyzing the non-superconducting limit of the model in momentum space, providing insight into the underlying band structure. Superconductivity is then incorporated via the Bogoliubov–de Gennes (BdG) formalism. Building on these momentum-space results, we next explore the system under open boundary conditions to examine edge phenomena. Finally, we introduce an antidot to investigate the emergence and structure of in-gap bound states. \\

\begin{figure*}[t]
\centering
\includegraphics[width=0.75\linewidth]{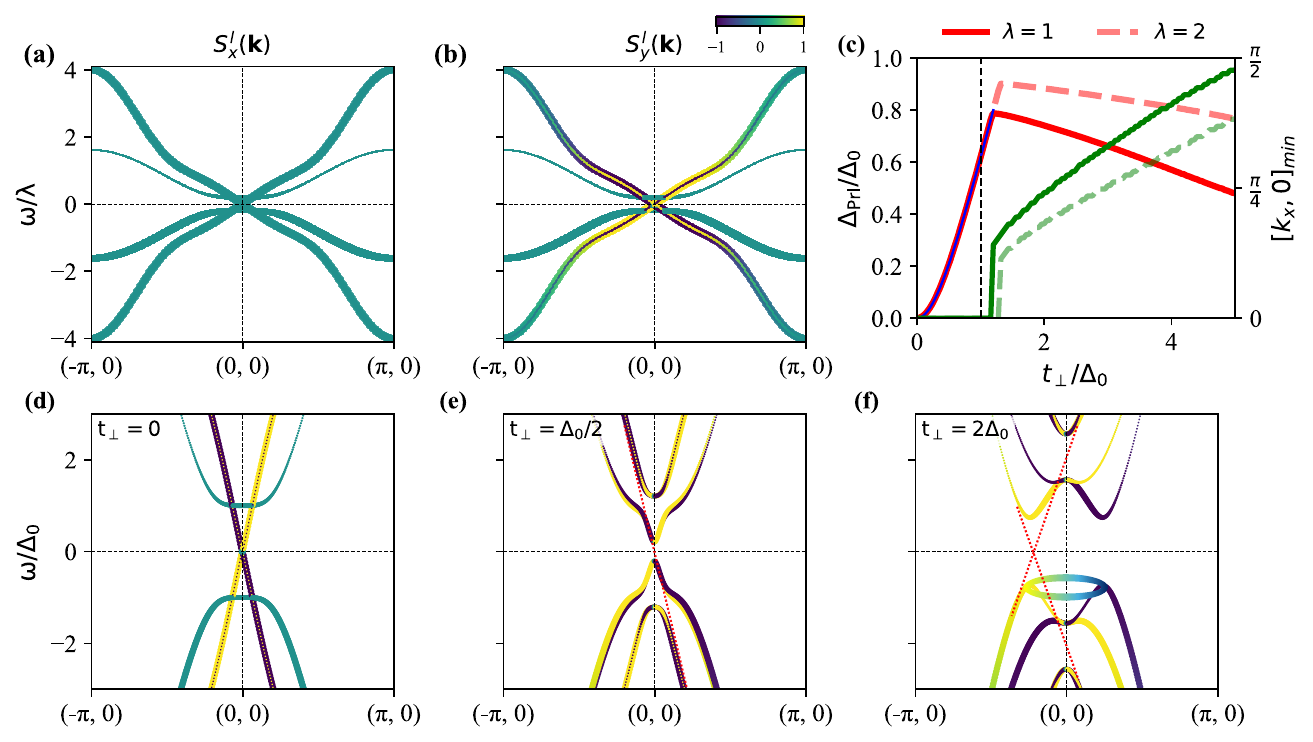}
\caption{\textbf{Gap opening in the SC--3DTI surface states bilayer.}  
(a) and (b) show the spin-resolved spectral functions with spin projections along the $x$ [$S_x^l(\bf k)$] and $y$ [$S_y^l(\bf k)$] directions, respectively, for $t_\perp = 0$, with the spin projections defined by Eq.~\ref{Eq:Sreslvd}. The thickness of the bands indicates the particle (thick) and hole (thin) components of the Bogoliubov quasiparticles.  
(c) Proximity-induced (PrI) gap as a function of interlayer tunneling strength $t_\perp$. Red lines represent numerical gap values. The blue line  given by Eq.~(\ref{eq:PrIGap_Gamma}) shows that the smallest gap opening at the $\Gamma$-point. Green line tracks the momentum-space location of the minimum gap in the Brillouin zone (BZ) for $\lambda = 1$ (the unit energy used throughout the text).  Faded lines in the panel show corresponding results for $\lambda = 2$ (equivalent to $\lambda=1$ with all other terms in the Hamiltonian halved).  
(d) Enlarged view of the low-energy spectral region from (b).   (e) and (f) display spectra for non-zero couplings $t_\perp = \Delta_0/2$ and $t_\perp = 2\Delta_0$, respectively, colored according to the spin projection in the $y$-direction. The dotted red lines in both panels  and serve as guides to the eye in the vicinity of induced gap. The projected circle (shown as an ellipse) in panel (f)  highlight that the band extrema form a circle in the BZ.  
}
\label{fig:bands}
\end{figure*}

%. without the superconductivity, i) 3DTI coupling and ii) metallic band

\textit{Metallic band–3DTI SS hybridization:} We now consider a non-interacting bilayer model composed of (i)  3DTI SS and (ii) a metallic band originating from the SC layer without a paring potential, $i.e.$, $\Delta =0$. The goal is to investigate how interlayer hybridization influences the band structure. The results are shown in figure~\ref{fig:Metal3DTI_int}. 

Figure~\ref{fig:Metal3DTI_int}~(a) displays the unhybridized bands ($t_{\perp} = 0$) of the two layers. The 3DTI SS exhibits spin–momentum locking near the Dirac point, as evident in the dispersion (despite the model explicitly breaking the TRS away from the $\Gamma$ point). By contrast, the metallic band from the SC layer is spin-degenerate and unpolarized.

Figure~\ref{fig:Metal3DTI_int}~(b) shows the hybridized band structure for a finite interlayer tunneling $t_\perp = \lambda/5$. At the crossing points, the spin-momentum-locked states of the 3DTI SS hybridize selectively with metallic states of the same spin, leading to the opening of spin-dependent hybridization gaps.  However, one spin-momentum-locked band remains always gapless  
resulting in only partial gapping of the spectrum, which leads to a Rashba-like dispersion of holes, 
consistent with inversion symmetry breaking. These features offer a sensitive probe of the interlayer coupling strength using spin-sensitive angle resolved photoemission spectroscopy~\cite{Moore2023}.

Away from the band crossing regions, the bands largely retain their character from the unhybridized limit. Importantly, the interlayer coupling enhances the 3DTI SS density of states near zero energy, compared to the uncoupled 3DTI SS, %\rrc{Compared to TI only? Overall the DOS seems smaller at 0 energy in Fig.~\ref{fig:Metal3DTI_int}(b) than in Fig.~\ref{fig:Metal3DTI_int}(a) due to the van Hove singularity in the SC layer.}, 
which can lower the energy of in-gap states. This contrasts with the single-layer 3DTI SS model described in Section~\ref{SingleLayer}, in which the density of states will be constrained by the TI spectrum alone. %\rrc{Should we make some introduction for the single-layer model?}

\subsection{Proximity induced Gap\label{sec:PrIGap}}
To explore the PrI gap in our model, we introduce a pairing potential $\Delta$ in the SC layer and solve the BdG Hamiltonian as presented in Eq.~(\ref{eq:BdGkspace}). Our focus is on the momentum-space structure of the 3DTI surface bands, the SC bands, and their hybridization, as illustrated in figure~\ref{fig:bands}.

Figures~\ref{fig:bands}~(a) and (b) show the spin-resolved band structure along the $\Gamma$--$X$ direction in the Brillouin zone (BZ), with spin projections computed using Eq.~(\ref{Eq:Sreslvd}) along the $x$- and $y$-axes, respectively. The bands span the range $[-4\lambda, 4\lambda]$ for $ m = 2\lambda$ used in our case. As discussed below Eq.~(\ref{eq:TIHamiltonian}),  large $m$ gaps out the Dirac points except for the one at the $\Gamma$-point and moves the band to $\pm 2 m$ at $(0,\pm \pi)$ and  $\pm 4 m$ at $(\pm \pi,\pm \pi)$   [see figure~\ref{fig:schematicsbilayer} (b)]  
Near zero energy along $\Gamma$-$X$, the spin is locked in the transverse $y$-direction,  while the $x$-component remains zero throughout the dispersion, consistent with the expected spin-momentum locking of the 3DTI SS. %\red{JV: This is a description of the presentation (not physics) and should be in the figure caption, not here:} 
%We have plotted both the thickness of the bands represents the particle (thick) and hole (thin) components of the Bogoliubov quasiparticles. 
Note that the particle and hole branches of the TI component, shown using thick and thin lines, respectively, are overlapping in figure~\ref{fig:bands}~(a).

The SC bands are gapped in the energy range $[-\Delta, \Delta]$, with dispersion in the interval $[\pm\Delta, \pm\sqrt{\Delta^2 + (4t_\text{SC})^2}]$ %\rrc{The square is outside the root. Should it be $(2 t_\text{SC})^2$?} \ok{I am confused. Since the full width of SC layer is $8t_{SC}$ without $\Delta$ and the band top is at zero energy, I think the correct formula is $\sqrt{\Delta^2 + (8 t_\text{SC})^2}$ for the band maximum measured at $(\pi,\pi)$. At $(\pi,0)$, this is $\sqrt{\Delta^2 + (4 t_\text{SC})^2} \sim 1.6 \lambda$}. 
These bands are spin-degenerate. Their form is chosen to qualitatively resemble experimental SC band structures reported in Ref.~\cite{doi:10.1126/sciadv.1602372}. Notably, the unhybridized bands preserve the TRS within the energy window of interest.

\begin{figure*}
\centering
\includegraphics[width=0.9\linewidth]{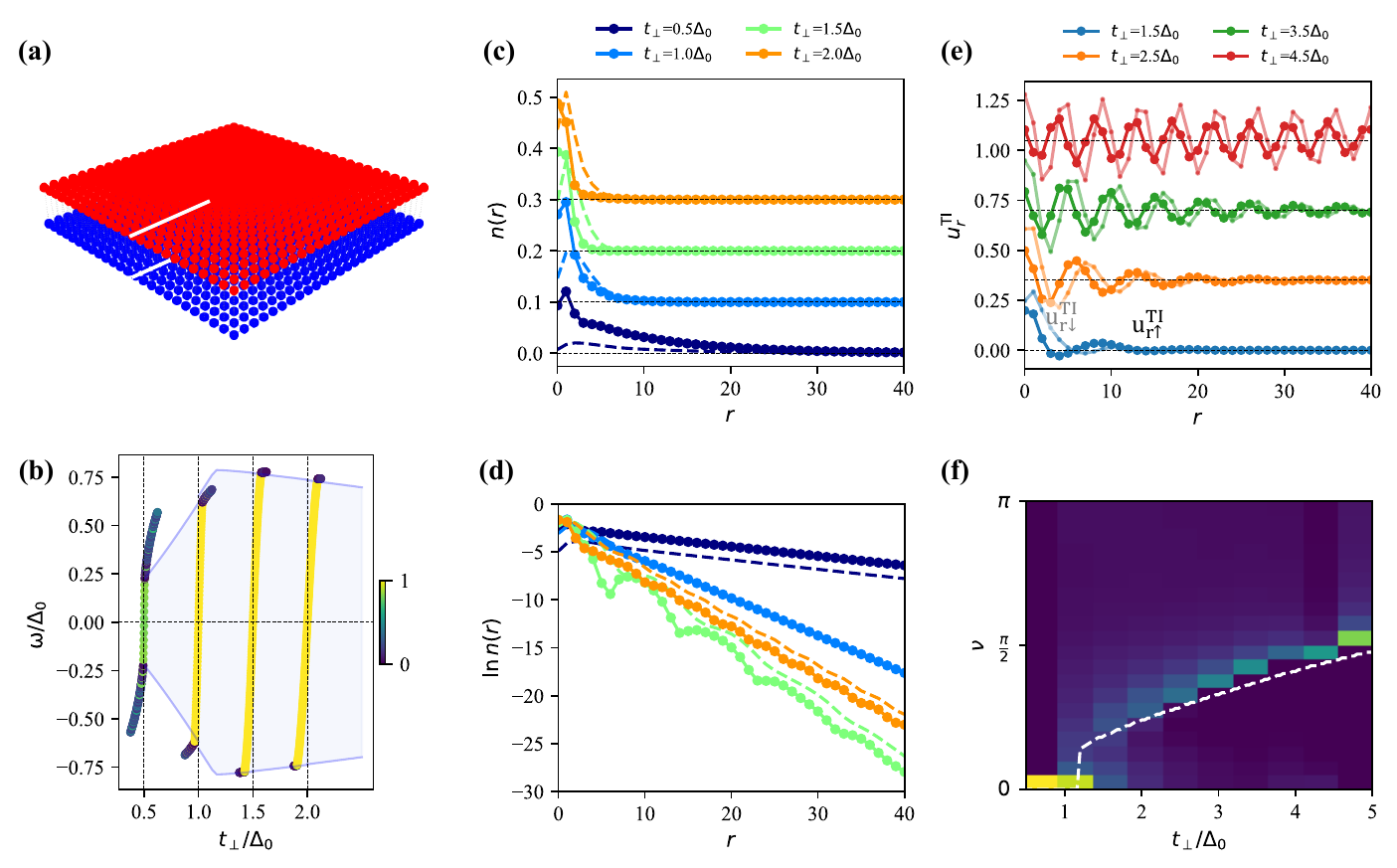}
\caption{\textbf{Chiral edge mode (CEM) in the bilayer model.}  
(a) Schematic of the SC--3DTI SS bilayer, with the spatial cut (white lines) used for analyzing edge-localized states.   (b) The 250 closest eigenvalues near the Fermi level for various interlayer couplings $t_\perp$. The color scale indicates the degree of edge localization in each eigenstate.  The eigenvalues are offset horizontally for better visibility. The blue line indicates the proximity-induced (PrI) gap  from figure~\ref{fig:bands}~(c) ($\lambda=1$), and the shaded region marks the corresponding gapped regime. (c) Spatial profile of the CEM probability density $n (r)$, shown separately for the TI $n_\text{TI} (r)$ and SC  $n_\text{SC} (r)$ layers using solid and dashed lines, respectively.   (d) Logarithmic plot of the probability density $n(r)$ from (c), highlighting the exponential decay into the bulk for both layers.   (e) Wavefunction components of the edge tail in the TI layer, showing particle spin-$\uparrow$,  ($u_{{\bf r}\uparrow}^\text{TI}$, solid lines) and spin-$\downarrow$ ($u_{{\bf r}\downarrow}^\text{TI}$ , faded lines) components. The amplitudes are rescaled by a decaying envelope to emphasize the oscillatory behavior.   (f) Oscillation frequency ($\nu$) of the CEM tail as a function of interlayer coupling $t_\perp$. The white line shows is the $[k_x, 0]_\text{min}$ values from figure~\ref{fig:bands}~(c)  ($\lambda=1)$. 
}
\label{fig:ChiralEdgeMode}
\end{figure*}

We now introduce interlayer coupling $t_\perp$ to investigate hybridization between the 3DTI surface states and the superconducting (SC) bands, and to analyze the resulting proximity-induced (PrI) gap in the hybridized spectrum [Figure~\ref{fig:bands}(c)]. The PrI gap is defined as the energy gap between the Fermi level and the nearest bulk states and is symmetric about the Fermi energy due to particle–hole symmetry.

Panel~(c) presents results for the 3DTI surface-state parameter $\lambda = 1$, which is used throughout the main text. For comparison, we also include results for $\lambda = 2$ (shown as faded lines), keeping all other parameters fixed, to illustrate how increasing $\lambda$ influences the PrI gap. The red points indicate the numerically extracted PrI gap at the $\Gamma$-point, while the blue curve shows the corresponding analytical estimate, derived below.

For small $t_\perp$, the PrI gap is located at the $\Gamma$ point [Figure~\ref{fig:bands}~(c)] and is given by
\begin{equation}\label{eq:PrIGap_Gamma}
    \Delta_\text{PrI} = \sqrt{ t_\perp^2 + \tfrac{|\Delta_0|^2}{2} - \tfrac{|\Delta_0|}{2} \sqrt{4 t_\perp^2 + |\Delta_0|^2}},
\end{equation}
which is also plotted as the blue curve in figure~\ref{fig:bands}~(c). In this regime, $\Delta_\text{PrI}$ is fully determined by $t_\perp$ and $\Delta_0$. In the limit $t_\perp / |\Delta_0| \ll 1$, the expression simplifies to $ \Delta_\text{PrI} \approx \frac{t_\perp^2}{\Delta_0}$, consistent with the expectation from second order perturbation theory. 

As $t_\perp$ increases beyond $1.2\Delta_0$ (specific to the model parameters used), the location of the PrI gap shifts from the $\Gamma$ point to a finite momentum, forming a ring-shaped contour of radius $|k_F|$ in momentum space [Figure~\ref{fig:bands}~(f)]. This transition is shown by the green line in figure~\ref{fig:bands}~(c). In this regime, the magnitude of the PrI gap remains below $\Delta_0$. The momentum-space transition is discontinuous: $|k_F|$ abruptly jumps from 0 to approximately $\pi/8$ at $t_\perp = 1.25\Delta_0$ (specific to the model parameters used), and then decreases monotonically with increasing $t_\perp$. This discontinuity in the gap location originates from the specific non-interacting metallic band structure assumed for the SC layer.

Figure~\ref{fig:bands}~(d) presents an enlarged view of the low-energy spectral region from figure~\ref{fig:bands}~(b) for the unhybridized case.  
Figures~\ref{fig:bands}~(e) and (f) illustrate the two characteristic regimes of PrI gap formation for interlayer couplings $t_\perp = \Delta_0/2$ and $t_\perp = 2\Delta_0$, respectively. For $t_\perp = \Delta_0/2$, the gap opens at the $\Gamma$ point.  
In contrast, for $t_\perp = 2\Delta_0$, the gap shifts to a non-zero momentum, forming a ring-shaped contour centered away from $\Gamma$.

%\red{JV: I would move this paragraph to be last since it's more forward-looking:}
Our work in the bilayer model here can be mapped to the single layer model explored in the literature, integrating out the SC layer~\cite{FuKane2008, Rafal2024}, the details of which are discussed in Sec.~\ref{SingleLayer}.  One peculiar thing in the single layer model, is that one need to tune the $\varepsilon_\text{TI}$ to get a gap minima at non-zero momentum in the bands. %\rrc{\ldots to get PrI gap minima at non-zero $|k_F|$.} %\rrc{I think we have not established by this point why comparisons to the single-layer model are worth considering (also, I don't think the single-layer model has been introduced). }

\subsection{Chiral edge modes  \label{sec:CEM}}
We now investigate the CEMs in the bilayer model by solving the real-space Hamiltonian with open boundary conditions using a $L=201$ square bilayer lattice, as described in Sec.~\ref{sec:Methods} and shown in figure~\ref{fig:ChiralEdgeMode}~(a).   
Coupling the 3DTI SS with an $s$-wave superconductor leads to a system with preserved TRS, hence the true Chern number of the hybrid system is $C=0$. However, in our effective Hamiltonian $C=-1$, due to the presence of the TRS-breaking term $M_\mathbf{k}$. By the bulk–boundary correspondence, this implies the presence of topologically protected edge states.

Notably, the edge of the system described by the effective Hamiltonian is topologically equivalent to the boundary between two regions of the 3DTI SS, one proximitized by a SC, and the other by a magnetic insulator. Consequently, the chiral edge modes investigated here can be interpreted as boundary modes of such a hybrid structure, in the limit where the induced gap on the magnetic-insulator-proximitized side becomes infinitely large   

Figure~\ref{fig:ChiralEdgeMode} summarizes our analysis of the CEMs.  
Figure~\ref{fig:ChiralEdgeMode}~(a) shows the schematic of the bilayer system, along with the spatial cut used in the analysis.  
Figure~\ref{fig:ChiralEdgeMode}~(b) presents the eigenenergies of the system as a function of interlayer hybridization strength $t_\perp$, shown along the horizontal axis.  
The color scale indicates the probability density localized at the edge of the lattice for each eigenstate: bright yellow denotes strong edge localization, while dark blue corresponds to bulk states.  
The edge-localized states form a gapless branch, consistent with the topological protection.  
 The bulk energy gap, highlighted by the pale blue background, agrees well with the PrI gap extracted from the momentum-space analysis in figure~\ref{fig:bands}~(c) for the corresponding values of $t_\perp$.

Interestingly, while the CEM states are nearly equally spaced in energy when the PrI gap minimum lies at the $\Gamma$ point (e.g., $t_\perp = \Delta_0/2, \Delta_0$), the levels become increasingly dense near the bulk continuum when the gap minimum shifts to a finite-momentum ring (e.g., at larger $t_\perp$). %\rrc{The signature of this\ldots (?)} 
The signature of which can be seen as the increase in the density of the states, ($i.e.~$ higher number of states in the same energy interval) shown close to the bulk for the large $t_\perp$ case.
%\red{JV: I don't like ``slope of states'' since the x-axis is not momentum. I would just write ``density of states''}
%\rrc{I can't really see the spacing, or tell the slope of the curves in Fig.~\ref{fig:ChiralEdgeMode}(b). We should either change the scale of the horizontal axis, or preface the statement above with something like "We've inspected thoroughly the spectra of edge states and found\ldots". }. 
This transition reflects a modification in the CEM's density of states across different hybridization regimes.

Figure~\ref{fig:ChiralEdgeMode}~(c) presents the spatial structure of the CEMs across both the SC and TI layers.  The probability density $n(r)$ is shown separately for the TI $n_\text{TI} (r)$  and SC $n_\text{SC}(r)$ layers using solid and dashed lines, respectively. %\rrc{At the end of section \ref{sec:Methods_vortex} symbols $ n_{TI}^m ({\bf r})$ and $ n_{SC}^m ({\bf r})$ are introduced. Is $n_r$ the same quantity as those two?  } 
The results are shown for a representative edge-localized state near the Fermi level along a spatial cut indicated by the white dashed line in figure~\ref{fig:ChiralEdgeMode}~(a).  
%Solid and dashed lines denote the TI and SC layers, respectively. 
Results are displayed for several values of $t_\perp$, annotated along the top of the panel.
%Figure~\ref{fig:ChiralEdgeMode}~(c) shows the spatial profile of the particle density for the chiral edge mode across the TI and SC layers.  
The TI-layer density, shown as solid lines, peaks sharply at the boundary and exhibits a rapid decay into the bulk.   As $t_\perp$ increases, the edge localization in the TI layer becomes more pronounced.  In the SC layer (dashed lines), the density is generally suppressed deep in the bulk.   A notable feature is the shift in the density maximum: at small $t_\perp$, the maximum is located away from the boundary but gradually shifts toward the edge as $t_\perp$ increases.  This reflects a redistribution of the edge mode's weight from the TI layer at small $t_\perp$ to the SC layer at larger $t_\perp$.

Moreover, the spatial profile in the SC layer at the edge evolves from Lorentzian-like at low $t_\perp$ to Gaussian-like at higher $t_\perp$, closely following the differences in the PrI gap transition from a minimum at the $\Gamma$ point to a finite-momentum contour (see panels (e) and (f) in figure~\ref{fig:bands}).  This evolution demonstrates that the SC-layer edge profile encodes key features of the underlying band structure and the nature of the induced gap.

To quantify the spatial decay of the edge mode, figure~\ref{fig:ChiralEdgeMode}~(d) plots $\ln n(r)$ for both layers over a range of $t_\perp$ values, where $n(r)~[=n_\mathrm{TI}(r)+n_\mathrm{SC}(r) ]$  %\rrc{Is $n(r)$ different from $n_r$ two paragraphs above?} 
denotes the spatially resolved edge-state density. %\red{JV: This should be in the figure caption, not in the text:}  
Solid and dashed lines represent results for the TI and SC layers, respectively.   The tails are modeled as exponential decays of the form $n(r) = A \mathrm{e}^{-r/\lambda_{\text{CEM}}}$, where $\lambda_{\text{CEM}}$ is the localization  length.  This leads to the following relation:
\begin{equation}
 \ln n(r) = -\frac{r}{\lambda_\text{CEM}} + \ln A
\end{equation}
The linear trends in $\ln (n(r))$ confirm this exponential behavior (slope $\propto 1/\lambda_\text{CEM} )$.   With increasing $t_\perp$, the slope becomes steeper, corresponding to the strong localization near the edge.  At large $t_\perp$, such as $t_\perp = 2\Delta_0$, the plots also display an oscillatory modulation superimposed on the decay, indicative of interference effects in the edge-state envelope.   Interestingly, the oscillations in the SC and TI layers are shifted, consistent with the spatial shift in the density maxima seen in figure~\ref{fig:ChiralEdgeMode}~(c).

%The inset shows the dependence of $\lambda_{\text{CEM}}$ on $t_\perp$, demonstrating that for large $t_\perp$, the edge states become highly localized—decaying over just a few lattice sites.

\begin{figure*}
    \centering
\includegraphics[width=\linewidth]{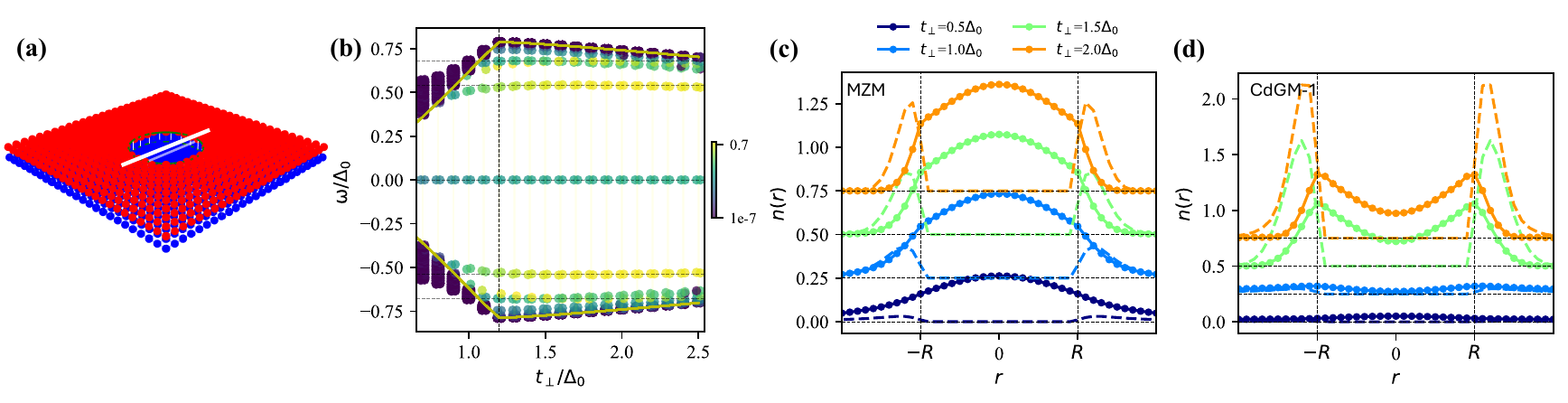}
\caption{\textbf{Majorana zero mode (MZM) and Caroli-de Gennes-Matricon (CdGM) states in the bilayer model.}  
(a) Schematic of the bilayer lattice with an antidot created in the superconducting layer. A magnetic vortex penetrates through the two layers in the antidot region.
(b) In-gap vortex-bound states for a set of interlayer couplings $t_\perp$. The color scale indicates the degree of localization in the vortex.  The yellow line plots an overlay of the proximity-induced (PrI) gap from figure~\ref{fig:bands}~(c) ($\lambda=1$). (c) Spatial profile of the probability density $n(r)$ for the vortex-localized MZM, shown along a linear cut through the vortex core in the 2D lattice depicted in (a). Solid and dashed lines represent the topological insulator (TI) and superconductor (SC) layers, respectively.  
(d) Same as (c), but for the first excited CdGM state.}
 \label{fig:Vortexmode}
\end{figure*}

We further investigate the spatial structure of the CEM in the strong interlayer coupling regime by analyzing the wave function components in the TI layer, for a representative edge state near $E = 0$, as shown in Figs.~\ref{fig:ChiralEdgeMode}~(e) and (f). Figure~\ref{fig:ChiralEdgeMode}~(e) displays the rescaled tails of  the wavefunction for several values of $t_\perp$, annotated above the panel. These are rescaled by an exponential factor given by $e^{+0.1r}$ to amplify the decay tail.  Solid and faded lines correspond to the spin-$\uparrow$ ($u^\mathrm{TI}_\mathbf{r\uparrow}$ )  and spin-$\downarrow$ ($u^\mathrm{TI}_\mathbf{r\downarrow}$) components, respectively. The wavefunction is rotated by a phase such that the $u^\mathrm{TI}_\mathbf{r\uparrow}$  has only real part and $u^\mathrm{TI}_\mathbf{r\downarrow}$ has only imaginary part, the magnitude of these is shown in the panel. A clear oscillatory pattern emerges in the spatial structure of the wavefunction of these edge-localized states, particularly pronounced in the strong-coupling regime.

To quantify the oscillation behavior, we perform a fast Fourier transform (FFT) on the spatial profiles and extract the dominant frequencies.  
Figure~\ref{fig:ChiralEdgeMode}~(f) plots these oscillation frequencies as a function of $t_\perp$ in the regime where the PrI gap minimum shifts away from the $\Gamma$ point in the Brillouin zone.  We observe a monotonic increase in the dominant oscillation frequency with $t_\perp$, consistent with the growing Fermi wavevector $k_F$ shown as the green line in figure~\ref{fig:bands}~(c).  This behavior further confirms that the spatial oscillations of the edge state are directly tied to the underlying momentum-space structure of the hybridized bands.
%These oscillations are most prominent for edge states near the PrI gap, though they are qualitatively present across all in-gap states \rrc{That's not shown in any figure.}.  
The broadening in frequency trend can be attributed to multiple PrI-gap minima across the Brillouin zone, leading to interference effects in the real-space profiles. %\rrc{What is the "monotonic frequency trend"?}

The oscillatory decay of the CEMs bears resemblance to the edge-state behavior in the Bernevig-Hughes-Zhang model of a 2D TI, where oscillations in the edge wave function emerge when the bulk band structure features a ring of extrema in $k$-space~\cite{QiZhang2011}. However, in our case, the presence of multiple bands near the Fermi level introduces additional complexity, leading to richer interference patterns and more intricate spatial structure.

% \rrc{ Oscillations in the topological states are discussed in \cite{Shan2010} for instance. There they consider suface states of a 3DTI and derive an algebraic condition for the oscillations to emerge, discussed above Eq. 15.
% This paper: \cite{Papaj2016} seems to discuss a similar thing in the first paragraphs of Sec. III. There are also mentions of the oscilation of the gap in thin films, resulting from the oscillation of the surface wave functions on opposite surfaces, like the supplement of \cite{Rechcinski2021} (see section S1 and Fig.~S1), \cite{Safaei2015,Liu2015}, but I don't think the wave-function oscillations are spelled out there. Also, I believe the edge wave-function oscillations can be reproduced also in a standard BHZ model, or in the $p$-wave SC model from \url{https://topocondmat.org/w7_defects/ti_majoranas.html}, with the right choice of parameters. All in all, I would not present this result as novel. }
% \ok{We can mention that such oscillatory behavior has been reported.}

%as its decays away from the edge and there are Freidel oscillations as minigap has finite $k$ at the minigap. The frequncy of the oscilattory part is estimated using $k = Im[ln()]$

\subsection{In-gap states bound to the vortex\label{sec:InGapStates}}
%\rrc{I would suggest a title like 'States bound to the antidot', as the edge states are also inside the PrI gap.} \uk{I meant the in-gap sates as these ar within the PrI gap and SC gap.} \rrc{That makes sense, but the edge states are also 'in-gap' in this sense, and are discussed in a separate section.}
Majorana zero modes can emerge in the core of a vortex that traps a single flux quantum in a topological $p$-wave superconductor.  
To model this scenario, we introduce a fabricated antidot of radius $R = 10$, where the SC lattice sites are removed within the vortex core, allowing it to trap a single flux quantum.  Details of this construction are provided in Sec.~\ref{sec:Methods_vortex}, and the geometry is illustrated in figure~\ref{fig:Vortexmode}~(a).  Inside the antidot, only the lattice sites of the 3DTI SS remain, and the flux is confined entirely within the vortex region of the topological layer. %\rrc{The change above is because in principle we could take into account the magnetic field penetrating into the SC part. We choose not to.}

Figure~\ref{fig:Vortexmode} presents the in-gap states localized at the vortex core in the bilayer model, highlighting two distinct classes: (i) MZMs and (ii) the CdGM states. 

Figure~\ref{fig:Vortexmode}~(b) shows the eigenvalue spectrum as a function of the interlayer coupling $t_\perp$ in the presence of a vortex. The color bar indicates the degree of localization within the vortex region: bright yellow corresponds to states strongly confined to the vortex core, while darker blue denotes more delocalized states. For clarity, edge-localized states have been removed from the plot. The yellow curve represents the bulk gap estimated from the momentum-space model, with the shaded region indicating the corresponding energy window. Notice that the eigenspectrum has a particle-hole symmetry across the zero energy mode.

We observe two zero-energy states for all the $t_\perp$ shown here, exhibiting strong and increasing localization in the vortex core, with the increasing $t_\perp$ indicated by the increasing weight of the yellow color. These states are identified as MZMs. The nature of these two MZMs; their internal structure including spin and spatial profiles is analyzed in subsequent sections where explicit forms of their wavefunctions are discussed.

%\rrc{In principle this coloring doesn't distinguish between the states in the antidot and delocalized states in the SC/TI layer. I suppose in practice, there are no delocalized states in the gap, but it would be more natural to color these dots according to the density in the antidot. I think in one of the meetings we discussed even excluding the edge states altogether.} \ok{I agree with Rafal}

Within the bulk gap we find several discrete states that are strongly localized in the vortex core. These states are identified as CdGM modes~\cite{CAROLI1964307}. The CdGM ``number” is assigned according to the energy separation from the MZM. 
Although they are topologically trivial, they differ qualitatively from CdGM states in a conventional $s$‑wave vortex; the contrasts are examined in detail in the wave‑function analysis section.

The first and second CdGM levels emerge at $t_\perp \simeq 0.8\Delta_0$.  The localization in vortex  increases with $t_\perp$ characterized by increasing yellow color, and the two levels remain nearly degenerate.  Notably, these modes energy rise with $t_\perp$ while the PrI gap grows, but remain essentially flat once the PrI gap begins to shrink.
A third CdGM branch appears at $t_\perp \simeq 0.9\Delta_0$, and their gap from the MZM appear to closely track the PrI gap.  Further, this mode localization is strongest when the PrI gap is largest, and diminishes as the PrI gap decreases. 
Beyond these, additional CdGM states are observed whose energies track the evolution of the PrI gap even more closely.

\begin{figure*}
    \centering
    \includegraphics[width=\linewidth]{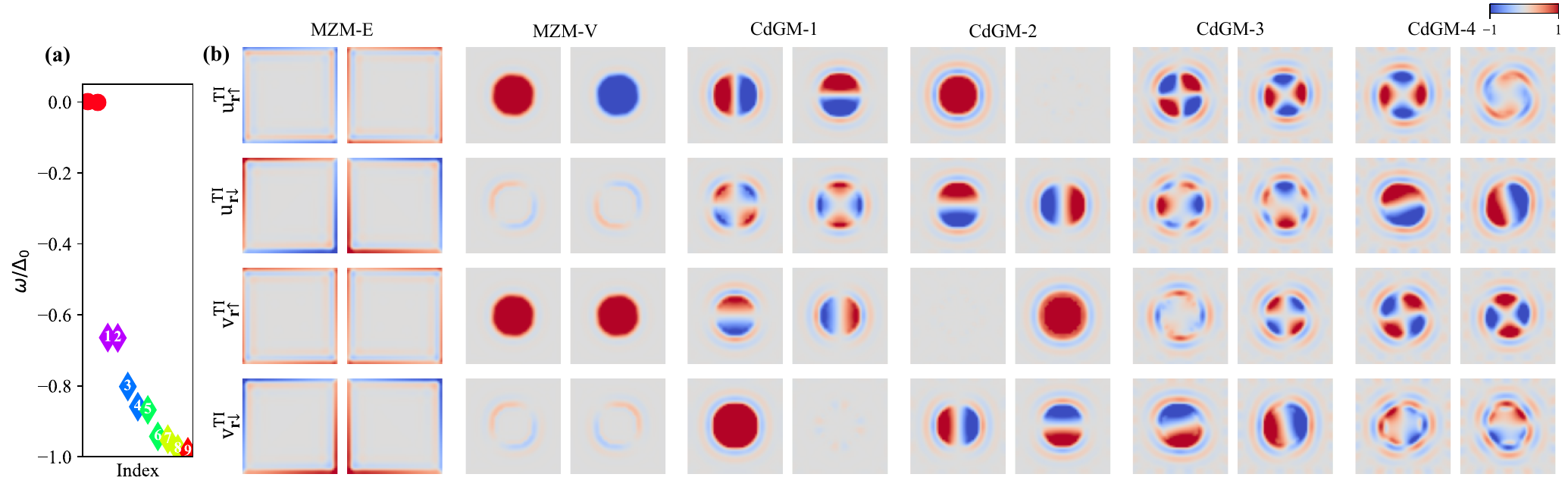}
    \caption{\textbf{Wavefunction components of vortex-bound states in the TI layer.}  
(a) Energy spectrum of vortex-bound states near the Fermi level.  
(b) Spatial profiles of the Bogoliubov--de Gennes wavefunction components $(u^{\text{TI}}_{{\bf r}\uparrow}, u^{\text{TI}}_{{\bf r}\downarrow}, v^{\text{TI}}_{{\bf r}\uparrow}, v^{\text{TI}}_{{\bf r}\downarrow})$ in the TI layer for the Majorana zero mode (MZM) and the four lowest-energy states adjacent to the MZM.    
Each pair of columns corresponds to a single eigenstate, with the left and right columns showing the real and imaginary parts of the wavefunction components, respectively. The color scale indicates the spatial amplitude of each component. Results are shown for interlayer coupling $t_\perp = 1.5\Delta_0$.
 } \label{fig:Wavefunction}
\end{figure*}

%We observe additional in-gap states, first mode of which is highlighted by the red dotted lines. These states can be identifies as the Caroli–de Gennes–Matricon (CdGM) states~\cite{CAROLI1964307}, although the boundary conditions here are somewhat different. We find that the CdGM states shift away from the Majorana zero mode (MZM) as the interlayer tunneling increases. This behavior can be attributed to the suppressed density of states from the superconducting (SC) layer near the vortex, due to the shorter decay lengths of in-gap states at higher hybridization. As a result, the particles bound in the vortex are primarily limited by the density of states of the topological insulator (TI) layer.

Figure~\ref{fig:Vortexmode}~(c) shows the probability density $n(r)$ of the MZM in the TI $n_\text{TI}(r)$ and SC $n_\text{SC}(r)$ layers, plotted using solid and dashed lines, respectively, along a spatial cut through the vortex core as indicated in panel (a).  In the TI layer, the MZM is sharply localized at the center of the vortex, with a pronounced maximum at its core.  In contrast, in the SC layer, the probability density peaks near the boundary of the antidot.  
Interestingly, the spectral weight in the SC layer increases more rapidly with interlayer coupling $t_\perp$ than in the TI layer, consistent with the trend observed for the CEMs, a feature of the enhanced hybridization with increasing $t_\perp$.

Figure~\ref{fig:Vortexmode}~(d) displays the probability density $n(r)$  of the first CdGM state for the same spatial cut, again distinguishing between the TI (solid line) and SC (dashed line) layers.  In the case of $t_\perp = 0.5 \Delta_0$, the CdGM state is not stabilized within the PrI gap, thefore the leading state can be understood as a bulk state as can be seen in figure~\ref{fig:Vortexmode}~(b). %\rrc{"This state has maximum density at the vortex center, similar to the MZM."  (to replace the next sentence, because 'bulk' implies no localization)} 
This bulk state has maxima near the vortex center. %\rrc{"However, as $t_\perp$ increases CdGM states appear within the PrI gap. The first CdGM  state, shown in the panel (d), is localized at the vortex boundary in both the TI and the SC layers." (fixed language; the last sentence seems to be repeating the previous one) }  
However, as $t_\perp$ increases the CdGM states appear within the PrI gap. The first CdGM (CdGM-1), shown in the panel, is localized at the vortex boundary in both the TI. In the SC layer too, there is a large weight localized at the outer edge of the antidot. 

The distinct spatial profiles of the MZM and first CdGM can be an important tool to distinguish between these two distinct modes relaized in the hybrid structure. We further delve into the precise nature of the wavefuction realized in the hybrid system below.\\

{\sl Wavefunction:---}
We now investigate the wavefunctions of the in-gap states near zero energy. The in-gap states are primarily localized in the TI layer, therefore, we focus on the TI part of the wavefunction given by  $| \Psi_n ({\bf r}) \rangle  \approx  (u_{\bf r \uparrow}^\text{TI} c_{\bf r \uparrow}^\dagger + u_{\bf r \downarrow}^\text{TI} c_{\bf r\downarrow}^\dagger +  v_{\bf r \uparrow}^\text{n} c_{\bf r \uparrow} + v_{\bf r\downarrow}^\text{TI} c_{\bf r \downarrow}) |0\rangle $, %\rrc{Is this notation standard? If $\gamma_{\bf r}^n$ is the creation operator, then it seems like it should be the sum of these vector components. } 
where $\bf r$ denotes the lattice site index within the TI layer.  Numerically, $\Psi_n ({\bf r})   = U^\dagger \phi_{\bf r}^\dagger $, here, $U^\dagger$ is the eigenvector of the real-space Hamiltonian, Eq.~(\ref{eq:BdGkspace}).
Since the wave function can be rotated by an arbitrary phase, we rotate the wave function by a global phase here, to plot the wavefunctions for better visualization. Figure~\ref{fig:Wavefunction} presents the energy spectrum and the corresponding site-resolved wavefunction components, computed for $t_\perp = 1.5\Delta_0$ and using a $L =51$ lattice size as it allows for better visualization of the wavefunction. Note that for this large $t_\perp$, the finite size effect is minimal. %\rrc{Would it make sense to also show the wave functions in the regime of $t_\perp>1.25 \Delta_0$ (where that gap is not at $\Gamma$)? } \uk{UK: The results don't change much except for a bit of Freidels oscillations enhancement.} 

Figure~\ref{fig:Wavefunction}~(a) shows the eigenenergies, %\rr{\st{for $E \le 0$, highlighting the MZM and the nearest CdGM states} of two MZMs and nearest CdGM states with negative energies} \rrc{one hybridized MZM shown in (a) has $E>0$} , 
and figure~\ref{fig:Wavefunction}~(b) shows the spatial wavefunction profiles of the two MZMs and selected CdGM states as indexed in panel (b). In each panel, the first and second columns correspond to the real and imaginary parts of the wavefunction, respectively. Note that the spectrum shown here has minor energy difference differences with the corresponding spectrum shown in figure~\ref{fig:Vortexmode}~(b), due to a reduced system size used here for better visualization, but does not affect the results otherwise.

%The spatial structure of the vortex MZM and CdGM states can be understood within the framework of a particle-in-a-disk potential. In this picture, the pairing potential acts as an effective infinite barrier, confining quasiparticles within the vortex region of the TI layer (see Appendix~\ref{App:FreeParticle} for details). This analogy provides a useful interpretation of the radial nodal patterns observed in the numerical wavefunctions. \rrc{I think the analogy with the particle in the disk is appropriate to the 2D SC, without the TI. Other than the fact that the solutions are eigenstates of angular momentum, there doesn't seem to be a lot in common between the TI/SC antidot and the 2DEG in a disk.   }

In the presence of a single vortex, the system hosts two degenerate MZMs. These modes can be re-orthogonalized into two spatially distinct solutions: (a) an edge MZM (MZM-E) and (b) a vortex MZM (MZM-V), as shown in panel (b). The edge MZM is entirely localized at the lattice's boundary edge, whereas the vortex MZM is confined within the vortex core.

Using the structure of the wavefunctions shown in figure~\ref{fig:Wavefunction}~(b), the two Majorana modes—localized at the vortex core and the system edge—can be expressed as:
\begin{subequations}
\begin{align}
\gamma_V = \sum_{\bf r} & \left[ \chi_V^\uparrow ({\bf r})\left( \mathrm{e}^{-i\frac{\pi}{4}} c_{{\bf r}\uparrow}^\dagger + \mathrm{e}^{+i\frac{\pi}{4}} c_{{\bf r}\uparrow} \right) \right. \notag \\
 +  & \left. \chi_V^\downarrow ({\bf r}) \left( -i Y_1^{+} c_{{\bf r}\downarrow}^\dagger + i Y_1^{-} c_{{\bf r}\downarrow} \right) \right] \label{eq:gammaV} \\
\gamma_E = \sum_{\bf r} & \left[ \chi_E^\uparrow ({\bf r}) \left( -i \mathrm{e}^{-i\frac{\pi}{4}} c_{{\bf r}\uparrow}^\dagger + i \mathrm{e}^{+i\frac{\pi}{4}} c_{{\bf r}\uparrow} \right) \right. \notag \\
 + & \left. \chi_E^\downarrow ({\bf r}) \left( Y_1^{+} c_{{\bf r}\downarrow}^\dagger + Y_1^{-} c_{{\bf r}\downarrow} \right) \right] \label{eq:gammaE}
\end{align}
\end{subequations}
where $ \chi_{V}^\sigma ({\bf r}) $ and $ \chi_{E}^\sigma({\bf r})$ denote the spin-resolved spatial envelopes for the vortex ($V$) and edge ($E$) Majorana modes, respectively and are real. The functions $Y_1^{\pm} = x \pm i y$ represent angular-momentum-based symmetry factors associated with the orbital structure of the states.
Note that the gauge choice for $\gamma_V$ is fixed to satisfy the Majorana condition, and the $\gamma_E$ has a relative $\mathrm{e}^{i\frac{\pi}{2}}$ phase, compared to the fermionic part shown in MZM-E wavefunction. In the given form, both the Majorana preserve the relation; $\gamma_{E(V)}^\dagger = \gamma_{E(V)}^{\phantom\dagger}$ and the anti-commutation relation; $\{\gamma_E, \gamma_V\} = 0$.

In the case of vortex MZM, we find that $|\chi_{V}^\uparrow| \gg |\chi_{V}^\downarrow|$, highlighting that it is almost fully spin-polarized along the direction of the magnetic field. The spin-$\downarrow$ component is negligible in magnitude and it shows an angular nodal structure that contrasts with the more nodeless spin-$\uparrow$ profile.  In contrast,  we find that for the edge Majorana, $|\chi_{E}^\uparrow| < |\chi_{E}^\downarrow|$, highlighting a relative spin flip at the edge mode. 

\begin{table}[t]
\begin{tabular}{|l|c|cccc|} \hline
$\qquad \qquad $ & $\quad m_J \quad$ & $u_\uparrow^\text{TI}$ &  $u_\downarrow^\text{TI}$ &  $v_\uparrow^\text{TI}$  &  $v_\downarrow^\text{TI}$    \\ \hline
MZM-V & $0$ & $0$ & $1$ & $0$ & $-1$  \\ \hline
CdGM-1 & $1$ & $1$ & $2$ & $1$ & $0$ \\
CdGM-2 & $0$ & $0$ & $1$ & $0$ & $-1$ \\
CdGM-3 & $2$ & $2$ & $3$ & $2$ & $1$ \\
CdGM-4 & $-2$ & $-2$ & $-1$ & $-2$ & $-3$ \\
\hline
\end{tabular}
\caption{Angular momentum quantum number $l$ associated with the wavefunctions of in-gap states in the 3DTI surface state/superconductor (SS/SC) hybrid vortex system,  as illustrated in figure~\ref{fig:Wavefunction}~(b).
}
\label{Table:MZMCdGM}
\end{table}

Figure~\ref{fig:Wavefunction}~(a) displays additional in-gap states distinct from the MZM and are numbered by integers increasing with energy relative to the MZM. Their corresponding wavefunctions plotted in the remaining columns of panel (b), up to the fourth CdGM state (CdGM-4). Since the states shown here lie below the Fermi level, their wavefunctions exhibit a dominant particle component, i.e., $|u_n| > |v_n|$, and a clear particle–hole asymmetry is evident in their spatial structure. Notably, all these CdGM states are localized within the vortex core and lack any corresponding edge state, indicating their topologically trivial nature. As a result, they can be fully described by the boundary conditions imposed by the vortex geometry.

The vortex setup considered here possesses rotational symmetry in two dimensions, implying that the total angular momentum operator $J_z$ commutes with the Hamiltonian in the continuum limit of the lattice. %\rrc{the lattice breaks the symmetry}. 
As a result, $J_z$ is a conserved quantity and can be used to label the eigenstates via a good quantum number, denoted by $m_J$. A variety of angular momentum symmetries consistent with the spatial structure of the numerically computed wavefunctions is summarized in Table~\ref{Table:MZMCdGM}. %\rrc{The ansatz follows from the Hamiltonian. We really shouldn't present it as motivated by numerical solutions. Perhaps one non-trivial fact is that the full ansatz should have 8 components (4 TI + 4 SC) and here we only show TI. However, the $(n/2)\tau_z$ term comes from the SC part.}  
We find that the wave functions shown in panel (b) are consistent with the ansatz:
\begin{equation}\label{eq:TIvortex}
 \Psi_{m_J} (r, \varphi) = \mathrm{e}^{i\{m_J+\frac{1}{2}(\sigma_z - N) \tau_z \} \varphi }   \Psi_{m_J} (r). 
\end{equation}
Here, $\sigma_z$ term accounts for the spin momentum locking in the two spin channels and the $N\in \mathbb{Z}$ denotes the vortex number [see Eq.~(\ref{Eq:SCGap})], while $\tau_z$ is the third Pauli matrix in particle-hole (Nambu) space. Furthermore, $m_J + (N/2)$ must be half-integer in order to ensure continuity of the wave function. The $J_z$ matrix in the TI subspace form can be written as: $\operatorname{diag}[m_J + (\frac{1}{2} -\frac{N}{2}), m_J + (-\frac{1}{2} -\frac{N}{2}), m_J - (\frac{1}{2} -\frac{N}{2}), m_J -  (-\frac{1}{2} -\frac{N}{2})]$. For $N =1$, $ J_z  = \operatorname{diag}[m_J, m_J-1,m_J,m_J+ 1)]$ with $m_J\in\mathbb{Z}$, is consistent with the results tabulated in Table~\ref{Table:MZMCdGM}. 
This ansatz is also consistent with the wavefunction reported for an antidot in a simpler model with the proximity-included implicitly in the TI SS Hamiltonian~\cite{deng2020bound,ziesen2021low} and induced topological superconductor in a superconducting nanowire shell~\cite{doi:10.1126/science.aav3392}. From the ansatz, is clear that the the 4 TI components in the wavefunction have a well defined angular symmetry, where $l_{u_\sigma}$ and $l_{v_\sigma}$ can account for the particle and hole sector, respectively in both the spin sectors.

In addition to the distinct angular momentum symmetries, the CdGM states exhibit several features that set them apart from the MZMs.

In particular, CdGM-2 shares the same angular quantum number as the MZM, with $m_J = 0$ in both cases. However, the CdGM-2 mode shows a significantly larger weight in the spin-$\downarrow$ sector, unlike the MZM. Additionally, a mild particle–hole asymmetry is observed, reflected in the unequal amplitudes of the particle ($u_n$) and hole ($v_n$) components.

CdGM-1 and CdGM-2, also identified in figure~\ref{fig:Vortexmode}~(b), appear nearly degenerate in energy. This quasi-degeneracy persists across the values of $t_\perp$ considered here, indicating that it is intrinsic rather than parameter-specific. These two modes are distinguished by their angular quantum numbers: $m_J = 1$ for CdGM-1 and $m_J = 0$ for CdGM-2, confirming their distinct character.

Interestingly, higher-energy modes such as CdGM-1, 3, and 4 correspond to increasing $|m_J|$, consistent with their interpretation as higher-order vortex excitations with $m_J = 1,2$ and 3, respectively. A defining feature of all these modes is difference in the angular nodes between the particle-hole sector for the spin-$\downarrow$ components, $i.e$, $|l_{u_\downarrow} -l_{v_\downarrow}| = 2$. 
%\red{JV: We have not defined $l$, please clarify what it means. Should it be replaced by $\Delta m_J$ ? }

In all the CdGM modes, there is an angular momentum shift between the spin-$\uparrow$ and spin-$\downarrow$ wavefunctions in both the particle and hole sector, attributable to spin-momentum locking in the vortex ($|l_{u_\uparrow(v_{\uparrow})}-l_{u_\downarrow(v_{\downarrow})}|=1$). 
%In contrast, no such angular shift is observed between the particle and hole sectors, indicating preserved particle–hole angular symmetry. %\rrc{Isn't there a shift in $\downarrow$ subspace?}

\begin{figure*}
\centering
\includegraphics[width=0.95\linewidth]{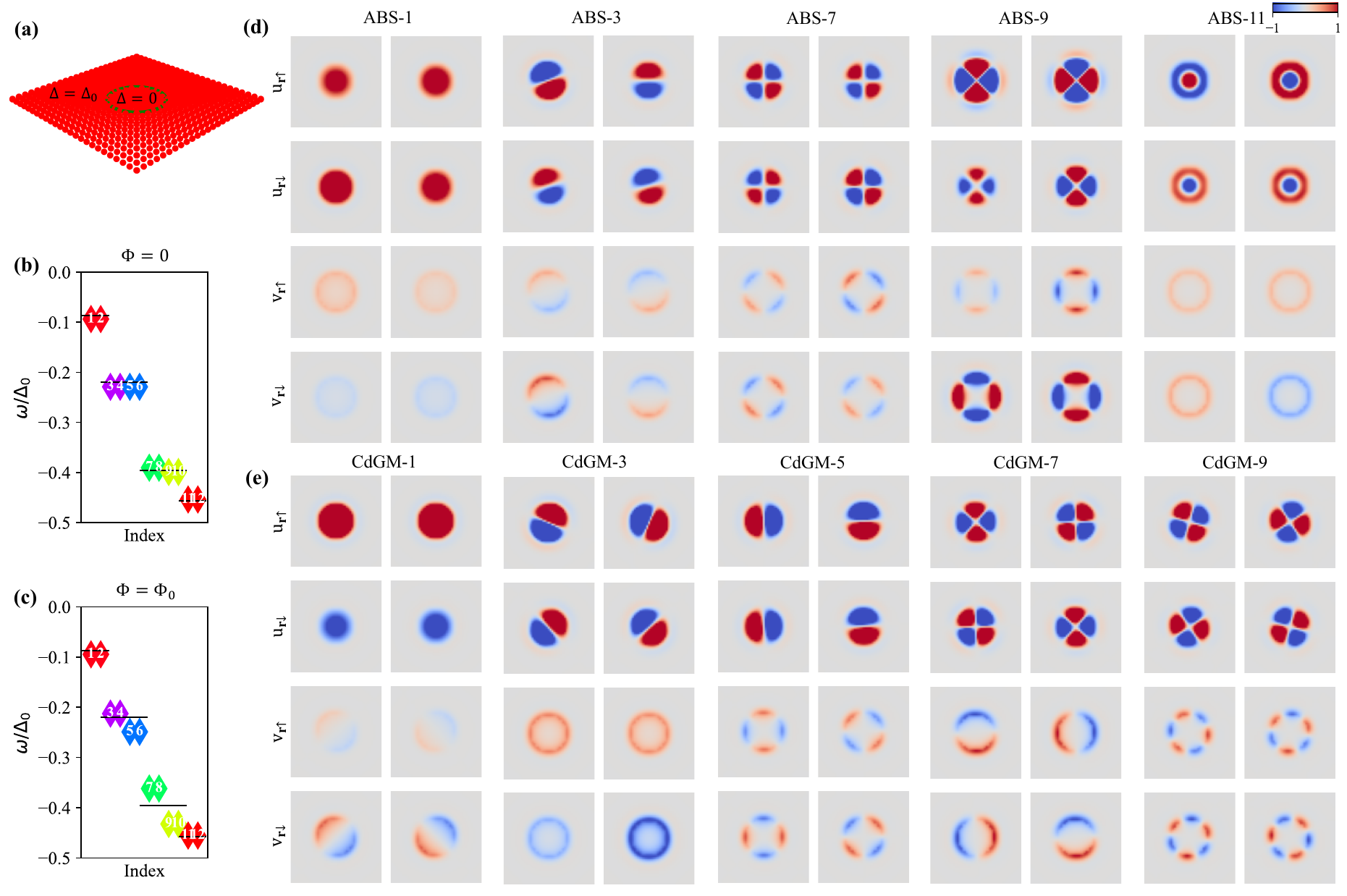}
\caption{\textbf{Vortex in an $s$-wave superconductor.}  
	(a) Lattice schematic of the $s$-wave superconductor with a vortex.  
	(b) and (c) In-gap energy spectra for the vortex configuration in the $s$-wave superconducting model without magnetic flux ($\Phi = 0$) and with magnetic flux ($\Phi = \Phi_0$), respectively. The black lines in both panels show the eigenenergies of free particles confined in a disk-shaped infinite potential well.  
	(d) and (e) Components of the Bogoliubov--de Gennes wavefunction $(u_{{\bf r}\uparrow}, u_{{\bf r}\downarrow}, v_{{\bf r}\uparrow}, v_{{\bf r}\downarrow})$ for vortex-bound states in the superconducting layer, corresponding to representative degenerate states from the spectra in panels (b) and (c), respectively.  
	For each state, the left and right subpanels show the real and imaginary parts of the wavefunction components. The color scale indicates the spatial amplitude of each component.}
\label{fig:Vortex_swave}
\end{figure*}

The intricate structure of the CdGM states stands in stark contrast to the in-gap states found in vortices of conventional $s$-wave superconductors, as will be further elucidated in the discussion on the $s$-wave case below. This richness arises from the spin-momentum locking inherent to the 3DTI surface, which plays a pivotal role in shaping the vortex-bound states. In particular, it is essential to the ansatz  in Eq.~(\ref{eq:TIvortex}) for capturing the symmetry structure of the wavefunctions.

Additionally, the in-gap states display Friedel-like oscillations in outside the vortex boundary, reminiscent of the oscillatory features observed in the chiral edge modes discussed in figure~\ref{fig:ChiralEdgeMode}. These oscillations become increasingly pronounced for in-gap states with energies approaching the bulk continuum, suggesting enhanced interference effects as the quasiparticle wavefunctions become less localized. \\

%Additionally, the particle component $u_n^\uparrow$, though weaker in amplitude, displays angular momentum characteristics with shifts of $\Delta l = {0, \pm 2}$ relative to $u_n^\downarrow$. This intricate angular momentum structure is a unique hallmark of CdGM states in s-wave SC coupled to the the 3DTI SS interface, distinguishing them from the simpler behavior observed in conventional $s$-wave superconductors with singlet pairing, as discussed in the following sections.

%\rrc{Are $u_n$ the hole or the particle components? Also, based on Fig.~\ref{fig:Wavefunction}(b) it seems that $\Delta l = \pm 1$ both between $u^\uparrow$ and $u^\downarrow$, and between $v^\uparrow$ and $v^\downarrow$. However, $\Delta l =0$ between $u^\uparrow$ and $v^\uparrow$, as well as $\Delta l = \pm 2$ between $u^\downarrow$ and $v^\downarrow$.  }
%\rrc{Refs.~\cite{deng2020bound} (Eq.~24) and~\cite{ziesen2021low} (Eq.~4) explain these values of $l$. }
% \\

{\sl Vortex in $s$-wave superconductor:---}
To better understand the nature of in-gap states in the TI/SC hybrid system and contrast them with those in a conventional $s$-wave superconductor, we analyze a simplified model consisting solely of the superconducting (SC) lattice described by Eq.~(\ref{eq:HamSC}), incorporating an antidot as illustrated in figure~\ref{fig:Vortex_swave}(a). Unlike the TI/SC hybrid system, in this case, the lattice sites are explicitly included inside the antidot region. The pairing potential is defined by Eq.~(\ref{Eq:SCGap}), with $\Delta = 0$ within the antidot—this discontinuity is the defining characteristic that distinguishes the antidot from the surrounding superconducting region.

We explore two cases for the model: (i) without a flux quantum ($\Phi = 0$) and (ii) with a flux quantum ($\Phi = \Phi_0$) trapped in the antidot marked by the green line in figure~\ref{fig:Vortex_swave}~(a). These setups allow us to explore the spatial dependence and structure of vortex-bound states. Figure~\ref{fig:Vortex_swave} presents the results for this analysis. Experimentally, several works have investigated vortex- and impurity-bound states in $s$-wave superconductors~\cite{PhysRevB.85.214505, Kong2021}, though a detailed comparison of the vortex state structure remains limited in the literature. Theoretical works comparing the vortex-bound states in the TI/SC hybrid to such states in a 2D SC layer have focused on Abrikosov vortices~\cite{deng2020bound} and on antidot-based vortices in the regime of $k_F\gg R^{-1}$~\cite{ziesen2021low}, distinct from the $k_F\approx 0$ regime considered here. 

Figure~\ref{fig:Vortex_swave}~(b) shows the energy spectrum of antidot-bound %\rrc{$\Phi = 0$ means no vortex} 
states in the lattice model for the zero-flux case ($\Phi = 0$), with eigenstates marked by diamond symbols and indexed by increasing energy relative to the Fermi level.

\begin{table*}[t]
\begin{tabular}{|c|c|cccc||c|c|cccc|} \hline
ABS & $\quad m_J \quad$ & $u_\uparrow$ &  $u_\downarrow$ &  $v_\uparrow$  &  $v_\downarrow$ &  CdGM  & $\quad m_J \quad$ & $u_\uparrow$ &  $u_\downarrow$ &  $v_\uparrow$  &  $v_\downarrow$  \\ \hline
ABS-1 & $1/2$ & $0$ & $0$ & $0$ & $0$ & CdGM-1 & $0$ & $0$ & $0$ & $-1$ & $-1$  \\ 
ABS-3 & $3/2$ & $1$ & $1$ & $1$ & $1$ & CdGM-3 & $1$ & $1$ & $1$ & $0$ & $0$ \\
ABS-7 & $5/2$ & $2$ & $2$ & $2$ & $2$ & CdGM-5 & $-1$ & $-1$ & $-1$ & $-2$ & $-2$\\
ABS-9 & $5/2$ & $2$ & $2$ & $2$ & $2$ & CdGM-7 & $2$ & $2$ & 2 & $1$ & $1$ \\
ABS-11 & $1/2$ & $0$ & $0$ & $0$ & $0$ & CdGM-7 & $-2$ & $-2$ & $-2$ & $-3$ & $-3$ \\
\hline
\end{tabular}
\caption{Angular momentum quantum number $l$ associated with the wavefunctions of in-gap states in a vortex hosted by an $s$-wave superconductor, shown for the cases with zero flux ($\Phi = 0$) and finite flux ($\Phi = \Phi_0$), as illustrated in figures~\ref{fig:Vortex_swave}~(d) and (e).}
\label{Table:swaveCdGM}
\end{table*}

%\rr{\st{$E_b \ll \Delta_0$} that the chemical potential lies at the band edge [see the metallic band in Fig.~\ref{fig:Metal3DTI_int}(a)] or in the forbidden region}. 

The spectrum exhibits clear signatures of quantization consistent with rotational symmetry. These states can be understood using an approximate continuum model describing a free particle of effective mass $m_b$ confined within a disk of radius $R$, where the confinement is provided by the superconducting gap $\Delta_0$, under the condition that the normal-state bands [see figure~\ref{fig:Metal3DTI_int}(a)] lie entirely within the superconducting gap, $i.e.$~$\omega_\text{SC}({\bf k} \rightarrow\Gamma) = |\mu-\varepsilon_\text{SC}-4t_\text{SC}|\ll|\Delta_0|$. 
Under this condition, the particle and hole branches in the Nambu picture do not overlap (or overlap in energy is much smaller than $\Delta$) and we can effectively describe the wave function by a single component. 
%\rrc{This condition is unclear, because the band spans from $-2t_{SC}$ to $0$. Also, the key condition is that the particle and hole branches in the Nambu picture do not overlap (or that their overlap is much smaller than $\Delta$). I suggest replacing the ending with "\ldots under the condition $|\mu|\ll|\Delta_0|$, i.e., that the Fermi level is near the edge of the normal-state band [see Fig.~\ref{fig:Metal3DTI_int}(a)]."  This ensures that if $\mu<0$, the overlap is small, and if $\mu>0$, then the band is not too far from zero energy (in that case $\mu \approx \sqrt{\Delta^2+\mu^2})$, so adding the pairing is  inconsequential). This is all under assumption that $\varepsilon_{SC}=-4t_{SC}$, otherwise we'd have to write $|\mu-\varepsilon_{SC}-4t_{SC}|\ll|\Delta_0|$ . } \rrc{I wrote the previous comment before noticing the condition $\omega_{SC}\ll\Delta_0$ in the appendix, which, after changing into absolute values, would be equivalent to $|\mu|\ll|\Delta_0|$. I prefer the version with $\mu$, as it doesn't require introducing $\omega_{SC}$.}

The quasiparticle states in this geometry satisfy a second-order differential equation, with solutions of the form $\Psi_b(r, \varphi) = \mathcal{R}(r)\, \mathrm{e}^{i l\varphi}$, where $l$ is the angular momentum quantum number. The solution to the equation for the radial wave function $\mathcal{R}(r)$ are Bessel functions $\mathcal{J}_l (kr)$. Imposing the boundary condition $\mathcal{J}_{l}(kR) = 0$ at the disk edge leads to a discrete set of allowed energy modes given by
\begin{equation}\label{eq:Ebnl}
E_b^{n l} = \frac{\hbar^2}{2m_b} \frac{\mathrm{x}_{n l}^2}{R^2},
\end{equation}
where $\mathrm{x}_{n l}$ is the $n^\text{th}$ zero of the Bessel function $\mathcal{J}_l (kr)$ (see Appendix~\ref{App:FreeParticle} for details). The levels given by Eq.~(\ref{eq:Ebnl}), shown as short black lines in panel (b), agree well with the lattice results using a fitted prefactor $\frac{\hbar^2}{(2m_b R^2)} = 0.015\Delta_0$ (see Appendix~\ref{App:FreeParticle} for details). In our lattice model in the limit $\mathbf{k} \rightarrow \Gamma$, the particle trapped in the antidot has the dispersion given by $|\omega_\text{SC}({\bf k}, \Delta_0 =0)| \approx t_\mathrm{SC} \mathbf{ |\mathbf{k}|}^2$. Using the relation $\hbar |\mathbf{k}|^2/(2m_b) = t_\mathrm{SC} |\mathbf{k}|^2$, this leads to to the  $\hbar^2/(2m_b R^2) = t_\text{SC}/R^2 \approx 0.02\Delta_0$, very close to the value used as prefactor above. The minor difference can be attributed to the finite SC gap used in the lattice model.
%\rrc{In our lattice model in the limit $\mathbf{k} \rightarrow \Gamma$ we have $E \approx -t_\mathrm{SC} (a\mathbf{ k})^2$ (including explicitly the lattice constant $a=1$). So, we should have $\hbar/(2m_b) = -t_\mathrm{SC}a^2$ and $\hbar/(2m_bR^2) = -t_\mathrm{SC}(a/R)^2$. How does the fitted prefactor compare to this?}

The solutions presented in figure~\ref{fig:Vortex_swave}(b) correspond to $\mathrm{x}_{10}$ (states 1 and 2), $\mathrm{x}_{11}$ (3, 4, 5, and 6), $\mathrm{x}_{12}$ (7, 8, 9, and 10), as well as $\mathrm{x}_{20}$ (11 and 12).
The red, magenta/blue, and green points correspond to angular momentum quantum numbers $l = {0, \pm 1, \pm 2}$ for the first radial mode ($n= 1$). %\red{JV: added $\pm$, please check that it's correct. }  %\rrc{Could the radial quantum number be relabeled $n_r$, as $n$ denotes the vortex number already?}. 
The $l = 0$ states are twofold degenerate due to spin, while $l \neq 0$ states are fourfold degenerate due to combined spin and orbital symmetries. Higher radial modes, such as the $n = 2$ states, also appear, with the subsequent red points representing $n = 2$, $l = 0$. %\rr{Adding the state indices besides the colors of associated points would make this paragraph easier to follow.}

Figure~\ref{fig:Vortex_swave}~(c) shows the energy spectrum of the vortex-bound states in the lattice model with finite magnetic flux ($\Phi = \Phi_0$), indicated by diamond symbols. For reference, the eigenspectrum of the corresponding particle in a disk continuum model is overlaid. In the presence of flux, the previously fourfold-degenerate states split into twofold-degenerate pairs. The $l = 0$ states retain their twofold degeneracy due to spin, while states with non-zero angular momentum ($l \neq 0$) exhibit reduced degeneracy; twofold—marking a clear departure from the fourfold degeneracy observed in the zero-flux case. This lifting of degeneracy originates from flux-induced splitting of the orbital angular momentum sectors: the magnetic field breaks time-reversal symmetry and couples directly to orbital, but not spin, degrees of freedom. The physical origin and implications of this behavior are further elaborated below.

%\uk{The energy states are plotted in panel (b) and (c) using the black lines as a reference to the eigenenergies of the lattice.} 
%\rrc{It would be useful to comment on the reasons for such good agrement with the simple Schr\"{o}dinger equation approach. It could be that it works only for $\hbar^2/(2 m R^2) \ll \Delta_0$ and $\mu \ll \Delta_0$, but this is based only on the fact that we seem to be in this regime. }

Figures~\ref{fig:Vortex_swave}~(d) and (e) show the corresponding wavefunctions from the lattice model for the vortex configurations with zero flux ($\Phi= 0$) and finite flux ($\Phi= \Phi_0$), respectively. Each column is labeled by the associated state number. As in previous figures, the left and right columns display the real and imaginary parts of the wavefunctions, respectively. %\rrc{Is there some procedure for choosing which wave function from the degenerate pair is plotted? The SC Hamiltonian separates into subspaces spanned by $(u^\uparrow, v^\downarrow)$ and $(u^\downarrow, v^\uparrow)$. One natural choice could be to show states confined to only one subspace. Fig.~\ref{fig:Vortex_swave} appears to show an unspecific superposition. } 

The rich angular symmetries evident in these panels and tabulated in Table~\ref{Table:swaveCdGM} are consistent with the ansatz for the vortex-bound wavefunctions~\rr{\cite{CAROLI1964307,deng2020bound,ziesen2021low}}. Incorporating the conserved total angular momentum quantum number $m_J$, the wavefunction can be expressed as:
\begin{equation} \label{eq:CdGMwavefunction}
\Psi_{m_J}(r, \varphi) = \mathrm{e}^{i \left\{ m_J + \frac{1}{2} (-\mathbb{I} - N) \tau_z \right\} \varphi} \, \Psi_{m_J}(r),
\end{equation}
where $\varphi$ is the polar angle, $\tau_z$ acts on the particle-hole space, and $N$ encodes the flux number. Similar to the TI/SC case, $m_J+(N/2)$ is half of an odd integer. This form captures the angular behavior consistent with the observed numerical solutions. Importantly in this case, spin symmetry is preserved ($|l_{u_\uparrow(v_{\uparrow})}-l_{u_\downarrow(v_{\downarrow})}| = 0$) in contrast to the spin-locking in the TI/SC interface case. %\red{leading to $\Delta l=\pm 1$ in the spin sector. 
%JV: I don't understand what this means... Which two wave functions differ in $l$ by $\pm 1$? Clarify the text. }

In the zero-flux case, the in-gap states can be interpreted as Andreev bound states (ABSs) confined within the normal region by the surrounding superconducting gap. These bound states are discrete solutions of a confined geometry, with wavefunctions shaped by the antidot boundary conditions.

Figure~\ref{fig:Vortex_swave}~(d) shows the spatial profiles of these wavefunctions, with each column labeled by the ABS index corresponding to its energy level. One representative state from each degenerate multiplet below the Fermi level is shown. The angular structure of these states is fully captured by the ansatz in Eq.~(\ref{eq:CdGMwavefunction}), which reflects the disk geometry and associated angular momentum classification.

Interestingly, the radial profiles reveal an asymmetry between particle and hole components: the particle sector resembles a Bessel function $\mathcal{J}_l$, while the hole sector appears shifted and resembles $\mathcal{J}_{l+ 1/2}$. This subtle difference results in radial phase shifts between $u_n$ and $v_n$. Nevertheless, no angular asymmetry is observed between the two sectors, consistent with preserved spin degeneracy in the absence of spin-orbit coupling. Additionally, the hole amplitude remains suppressed relative to the particle component, as expected from the bound-state condition $|u_n| > |v_n|$ for states below the Fermi level.

%The energy spectrum of the in-gap states for $\Phi = 0$ is shown in Fig.~\ref{fig:Vortex_swave}~(b). \uk{The states in these case are the ABS, trapped in vortex by the superconducting gap outside the vortex.}  
%\rrc{Since there’s no phase winding, referring to these as CdGM states (even with this disclaimer) could be misleading to readers. Also, this is the only place in the paper where this generalization/redefinition of the terminology would be relevant, so it's probably not worth the potential confusion. If you prefer the plot titles to be unified, instead of "CdGM" perhaps we could simply use "wave function" or "bound state" in in Fig.~\ref{fig:Vortex_swave}(d,e) (and possibly also Fig.~\ref{fig:Wavefunction}(b), maybe with some additional label for the MZMs).}

We next examine how the vortex-bound states change upon introducing a single magnetic flux quantum ($\Phi = \Phi_0$), where the in-gap states can be understood as Caroli-de Gennes Matricon (CdGM) states, for which the spectrum is shown in panel (C). The corresponding energy wavefunctions are shown in panel~(e)  with the label for the CdGM indexed on the each column. Again, Eq.~(\ref{eq:CdGMwavefunction}) fully captures the angular symmetry realized in the continuum limit of the lattice. The particle and hole sector radial dependence show the behavior akin to zero flux case. However, a key difference emerges in the angular behavior: a mismatch of $\Delta l = 1$ is observed between the particle and hole sectors, reflecting the phase winding introduced by the magnetic field in the vortex. The spin-resolved wavefunctions remain aligned ($\Delta l = 0$), preserving the spin symmetry, as the spin doesn't directly couple to the flux. Here too, states below the Fermi level continue to exhibit dominant particle components ($|u_n| > |v_n|$).

%As these states lie below the Fermi level, their wavefunctions are dominated by the particle component, i.e., $|u_n| > |v_n|$. Notably, we observe an angular momentum shift of $\Delta l = 1$ between the particle and hole components. However, the spin-resolved wavefunctions exhibit no angular shift ($\Delta l = 0$) between different spin sectors. This behavior is in contrast to the bilayer system, where spin-sector shifts of $\Delta l = \pm 1$ were observed due to effective triplet pairing and interlayer hybridization.

These results highlight the qualitatively distinct nature of in-gap states in conventional $s$-wave superconductors compared to systems exhibiting emergent $p$-wave pairing. Our spatial and spectral analysis underscores the unique fingerprints of topological superconductivity in engineered heterostructures, providing a robust framework for interpreting future experiments on vortex-bound states and unconventional pairing mechanisms.
%\rrc{In Ref.~\cite{ziesen2021low} there's a similar comparison between vortex-bound states  in TI/SC and 2DEG/SC hybrids.}  %\rrc{Also Ref.~\cite{deng2020bound}, but with different gap profile. }

\subsection{Single layer  model: SC layer integrated out}\label{SingleLayer}
\begin{figure}
\centering
\includegraphics[width=\linewidth]{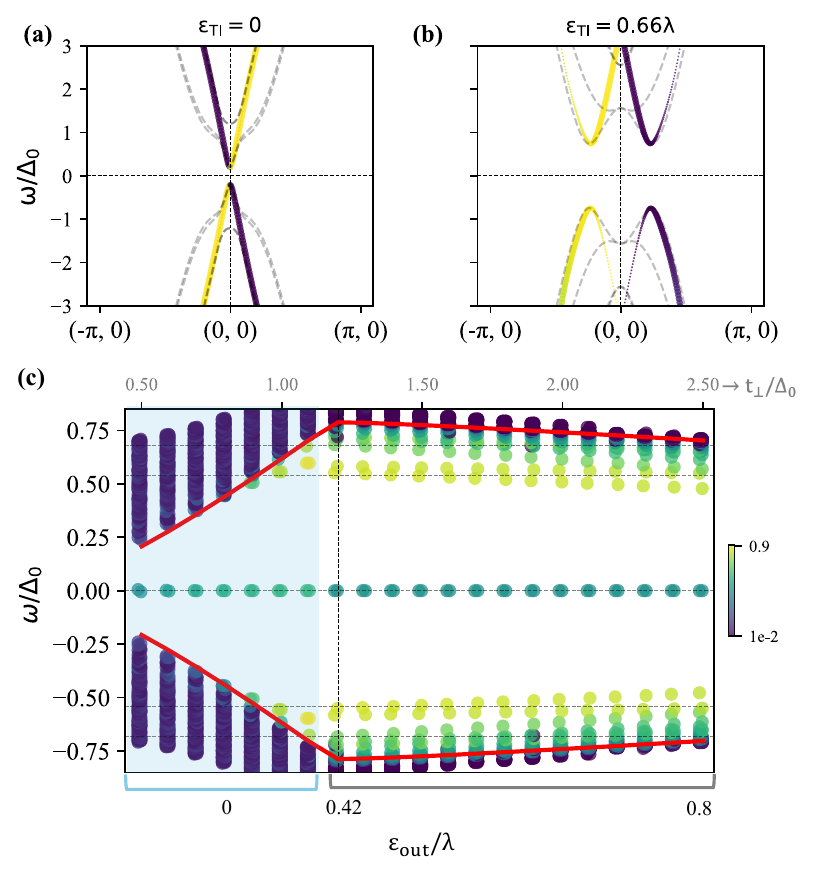}
%\caption{Single layer model with the SC layer is integrated out. a) and b) shows the band structure for $\varepsilon_\text{TI} =0$ and $\varepsilon_\text{TI} =0.7\lambda$ for $\Delta_\text{eff} =0.2\Delta_0$ and $0.75\Delta_0$ respectively. The color yellow (blue) color indicate the spin polarization parallel (antiparallel) to $y$-direction. The dashed black lines are an overlay for the respective bilayer model. c) shows the energy states for the gap at $\Gamma$  ($\varepsilon_\text{TI} = 0$ and $\Delta_\text{eff} = \{0.4\Delta_0, 0.8\Delta_0\}$), and for SC gap away from $\Gamma$ ($\varepsilon_\text{TI} = 0.5\lambda$, $\Delta_\text{eff}=0.8\Delta_0$). The yellow color indicate the weight at the edge of the lattice. The light thin points are the results from the bilayer model for $t_\perp$ indicated on the $x$-axis. }
\caption{\textbf{Single layer model}. a) and b) shows the band structure for $\varepsilon_\text{TI} =0$ and $\varepsilon_\text{TI} =0.66\lambda$ for $\Delta_\text{eff} =0.2\Delta_0$ and $0.74\Delta_0$ respectively. The yellow (blue) color indicates the spin polarization parallel (antiparallel) to $y$-direction. The dashed black lines are an overlay for the respective bilayer model. c) shows the energy states for the in-gap states evaluated using a single layer model with an antidot trapping a single magnetic flux quantum, for comparison with figure~\ref{fig:Vortexmode} (b). Here the chiral edge states are removed and the red line is the PrI gap ($\Delta_\text{PrI}$) and values on top of the panel are the corresponding $t_\perp/\Delta_0$ values from the bilayer model. The shaded blue region is for that states where the band minima in $\mathbf{k}$-space are at $\Gamma$-point, where the remaining data shows for the minima away from $\Gamma$, simulated by changing $\varepsilon_\text{TI}$ outside the vortex ($\varepsilon_\text{out}$), indicated below the panel.  }
\label{fig:singlelayer}
\end{figure}

%\rrc{Since this model is refered to in the previous sections, should its description be moved to the Methods section?} \uk{UK: I believe it is alright here. $\Delta_{eff}$ is used only in this section}
Here, we discuss routes to use an effective single-layer model~\cite{PhysRevX.7.031006,Rafal2024}, in which the SC layer is integrated out. The Hamiltonian is given by
%\rrc{I added the Hamiltonian, because $\Delta_\mathrm{eff}$ was used further down without any explanation. }
\begin{equation}
    H_\mathrm{eff} = H_\mathrm{TI} + \sum_{\bf r} \bigl( \Delta_{\mathrm{eff,}\bf r} \psi_{\bf r}^\dagger  \sigma_y  \psi_{\bf r}^\dagger + \text{h.c.}\bigr)
\end{equation}
%\rr{with the SC layer integrated out \st{by integrating out the SC layer}}, 
Here $H_\mathrm{TI}$ is the 3DTI SS Hamiltonian given by Eq.~(\ref{eq:TIHamiltonian}) and $\Delta_{\text{eff}, {\bf r}} = \Delta_\text{eff}$ is the effective pairing used in the model. Same as in the bilayer model, we use $\Delta_\text{eff} = 0$ inside the antidot and a finite $\Delta_\text{eff}$ outside the vortex, given by Eq.~(\ref{Eq:SCGap}). The $\Delta_\text{eff} = \Delta_\text{PrI}$ is used to reproduce the characteristics of the bilayer model discussed above. The results of the effective single layer model are shown in figure~\ref{fig:singlelayer}. %\rr{\st{One has to use a piece-wise limit on the bilayer model to reproduce its characteristics in the single layer, the details for which are discussed below:} 
Depending on the location of the band extrema in the bilayer model, a different corresponding parameterization scheme of the single-layer model is necessary. To achieve the agreement between the two models, $\varepsilon_\mathrm{TI}$ may attain different values inside the antidot ($\varepsilon_\text{in}$) and outside of it ($\varepsilon_\text{out}$). \\

(a) {\sl Minimum at $\Gamma$:---} In the single-layer model, setting $\varepsilon_\text{in/out} = 0$ ensures that the band minimum remains fixed at the $\Gamma$ point. This behavior mimics the bilayer model in the regime of low interlayer tunneling strength ($t_\perp$), where the proximity-induced (PrI) gap also exhibits its minimum at $\Gamma$, as shown in figure~\ref{fig:bands}(c). Since the PrI gap in the bilayer model depends on $t_\perp$, we incorporate this effect into the single-layer model by replacing the intrinsic pairing potential with an effective value, $\Delta_\text{eff} \equiv \Delta_\text{PrI}$. Panel~(a) compares the band structure obtained from the single-layer model (thick lines) using $\Delta_\text{eff} = 0.2\Delta_0$ to the bilayer result (black dashed line) at $t_\perp = 0.5\Delta_0$, which yields $\Delta_\text{PrI} = 0.2\Delta_0$. The good agreement confirms the validity of this effective model approximation.

Panel~(c) presents the spectrum of in-gap states in the single-layer model with an antidot trapping a magnetic flux. The shaded blue region highlights results where the band minimum is constrained to the $\Gamma$ point by setting $\varepsilon_\text{in/out} = 0$, while the effective pairing $\Delta_\text{eff}$ is matched to the PrI gap extracted from the bilayer model. The red line indicates the PrI gap as a function of $t_\perp$ from the bilayer calculation. The horizontal axis corresponds to different values of $t_\perp/\Delta_0$ ranging from 0.5 to 1.1. This simplified model reproduces the MZM--CdGM energy gap observed in the full bilayer calculation [see figure~\ref{fig:Vortexmode}(b)] with remarkable accuracy.

(b) {\sl Minima at finite ${\bf k}$:---} A band structure with minima at finite ${\bf k}$ in the Brillouin zone can be realized in the single-layer model by tuning the on-site energy $\varepsilon$ and adjusting it to match the finite-${\bf k}$ minimum observed in the bilayer model, as shown in figure~\ref{fig:bands}(c). To reproduce the proximity-induced (PrI) gap, we again set $\Delta_\text{eff} = \Delta_\text{PrI}$. Panel~(b) displays the resulting band structure from the single-layer model (thick lines), using $\Delta_\text{eff} = \Delta_\text{PrI} = 0.74\Delta_0$ and $\varepsilon_\text{TI} = 0.66\lambda$, %\rrc{$\varepsilon_\text{TI} = 0.66\lambda$ (since this is without the antidot)}, 
chosen to simulate the bilayer model with $t_\perp = 2\Delta_0$ (black dashed line). The excellent agreement at low energies confirms the efficacy of this mapping.

Panel~(c), outside the blue-shaded region, shows the in-gap states in the single-layer model for a vortex trapping a magnetic flux. These states are computed using $\varepsilon_\text{in} = 0$ and $\varepsilon_\text{out}$ varied from $0.42\lambda$ to $0.8\lambda$, as labeled along the horizontal axis. This choice reflects the physical setting of the bilayer model, where the 3DTI SS layer inside the antidot remains unaffected due to the absence of the superconducting layer in that region. This mapping captures the bilayer behavior for $t_\perp/\Delta_0$ ranging from 1.2 to 2.5.

Overall, the single-layer model reproduces the in-gap spectra of the bilayer system with good fidelity, though some differences emerge. In particular, for $t_\perp = 1.2\Delta_0$, a mild discontinuity is observed compared to the bilayer case. Additionally, the lowest CdGM mode shifts toward the MZM as $\varepsilon_\text{out}$ increases (i.e., as the PrI gap decreases), in contrast to the bilayer case where this mode remains nearly constant.
.
%This is in spite of the considering a different $\varepsilon_{TI}$ in the vortex, otherwise, its becomes negligible. %Further, one can get Freidels oscillations using this model as realized in the bilayer model. 

\section{Conclusions\label{conclusions}}

In this work, we have developed and analyzed a bilayer model to study the interface between the surface states of a three-dimensional topological insulator (3DTI) and  a 2D $s$-wave superconductor. By systematically varying the interlayer tunneling strength ($t_\perp$), we investigated  evolution of the proximity-induced superconducting gap, the formation of chiral edge modes (CEMs), and the emergence of in-gap states, including Majorana zero modes (MZMs) and Caroli–de Gennes–Matricon (CdGM) states. %\rrc{these abbreviations are introduced already}

Our results demonstrate that interlayer hybridization plays a central role in modifying the proximity-induced (PrI) gap 
and spectrum at the Fermi level. As $t_\perp$ increases, the PrI gap shifts away from the $\Gamma$-point, leading to a finite Fermi momentum. This induces Friedel-like oscillations in the spatial dependence of the in-gap states near defects and boundaries.
%\rr{\st{These oscillations are reflected in both the momentum-space structure and spatial decay profiles of the edge states.}}\rrc{This just repeated the previous sentence.} 
Notably, the superconducting layer contributes nontrivially to the edge-state localization and spin structure, particularly in the strong-coupling regime.

By introducing an antidot to emulate a vortex, we examined the resulting in-gap spectrum and identified robust MZMs localized at both the system boundary and vortex core. 
The energy spacing between MZMs and CdGM states is significantly enhanced at large $t_\perp$, driven by suppressed density of states in the SC layer near the antidot. Spin-resolved wavefunction analysis further reveals angular momentum shifts in the CdGM states—absent in conventional $s$-wave systems—highlighting the emergent $p$-wave character of the bilayer system. A direct comparison with a standalone $s$-wave superconductor confirms that in-gap states in the bilayer exhibit qualitatively distinct spatial and spectral signatures due to the interplay of topology and proximity effects.

We further introduced an effective single-layer mapping that captures key features of the bilayer system, including the vortex spectrum and gap evolution, while substantially reducing computational complexity. This effective model can serve as a valuable tool for qualitative exploration of vortex-bound physics.   

Together, these findings offer a microscopic understanding of the role of interlayer coupling in topological SC–3DTI heterostructures and provide experimentally relevant predictions for tuning and identifying Majorana modes. This work establishes a foundation for future explorations of topological superconductivity in engineered interfaces, including investigations of disorder effects, braiding protocols, and the impact of additional symmetry-breaking perturbations.

Our work is also relevant to further investigate the role of a ``topological proximity effect.''  As reported in Bi/TlBiS$_2$ heterostructures~\cite{Shoman2015},  the surface bands of a 3DTI can migrate to an attached metallic layer.  This effect could also lead to the topological superconductivity when the attached metal undergoes a superconducting transition.  In principle, both the superconducting proximity effect and the topological proximity effect should exist in any hybrid structures consisting of a topological insulator and a superconductor, and the relative strength between the two proximity effects is controlled by the hybridization strength between the two constituent systems as we studied in this work.  Future work considering realistic materials parameters will provide further insight into the nature of proximity-induced superconductivity. 

\begin{acknowledgments}
This work has been supported by the U.S. Department of Energy, Office of Science, National Quantum Information Sciences Research Centers, Quantum Science Center. We thank R. G. Moore, Micha\l{} Papaj and Eugene Dumitrescu for their useful discussions. This research used resources of the Compute and Data Environment for Science (CADES) at the Oak Ridge National Laboratory, which is supported by the Office of Science of the U.S. Department of Energy under Contract No. DE-AC05-00OR22725.
\end{acknowledgments}

\appendix

%\section{Appendixes}
%\onecolumngrid

%\ok{\bf Appendices should rearranged as they are referenced in the main text.}

\section{Phase factors\label{sec:PhaseFactors}}%Details on the Gauge Transformation}
In the presence of a vector potential, both electron hopping and the SC paring potential acquire phase factors. 
Since different notations have been used in literature, here we summarize our choice of phase factors based on the gauge transformation.

To incorporate the effect of magnetic flux, the vector potential $\mathbf{A}$ must satisfy the constraint $\oint \mathbf{A} \cdot d\mathbf{l} = \frac{\hbar}{e} \Phi$, where $\Phi$ is the enclosed flux. Here we have set speed of light $c=1$.  This results in the emergence of a Peierls phase in the hopping term, such that %$c_i^\dagger c_j \rightarrow \mathrm{e}^{i\varphi_{ij}} c_i^\dagger c_j$, where $\varphi_{ij} = \frac{e}{\hbar c} \int_{\mathbf{r}_j}^{\mathbf{r}_i} \mathbf{A} \cdot d\mathbf{l}$.
$c_{\bf r}^\dagger c_{\bf r'} \rightarrow \mathrm{e}^{i\Phi_{\bf r r'}} c_{\bf r}^\dagger c_{\bf r'}$, where $\Phi_{\bf r r'} = -\frac{|e|}{\hbar} \int_{\mathbf{r}}^{\mathbf{r}'} \mathbf{A}({\bf s}) \cdot d\mathbf{s}$.
%\ok{Here, we should use the same notations as we used near Eq.(7).}

Under a local $\mathrm{U}(1)$ gauge transformation characterized by a scalar field $\chi(\mathbf{r})$, the vector potential transforms as $\mathbf{A}'(\mathbf{r}) = \mathbf{A}(\mathbf{r}) + \nabla \chi(\mathbf{r})$. Correspondingly, the fermionic and pairing fields transform as:

\begin{align}
    & c_{\bf r} \rightarrow \mathrm{e}^{i\chi (\bf r)} c_{\bf r}, \\
    & \Delta_{\bf r} \rightarrow \mathrm{e}^{2i\chi({\bf r})} \Delta_{\bf r}.
\end{align}

Under these transformations, the Bogoliubov–de Gennes (BdG) Hamiltonian remains gauge invariant.

Importantly, the total flux through the system must be zero outside the vortex~\cite{degennes1999superconductivity}. This condition requires that the transformed vector potential satisfies $\oint \mathbf{A}'(\mathbf{r}) \cdot d\mathbf{l} = 0$ when integrated around the entire system. Applying this to the gauge-transformed vector potential leads to:
\begin{align}
    \oint \nabla \chi(\mathbf{r}) \cdot d\mathbf{l} = - \oint \mathbf{A} \cdot d\mathbf{l}.
\end{align}
This implies a specific structure for the scalar field:
\begin{align}
    \chi(r, \varphi) = -N \varphi + \chi_0,
\end{align}
where $\chi(r, \varphi)$ is independent of $r$ and varies only with the polar angle $\varphi$. The winding number $N$ corresponds to the number of flux quanta trapped in the vortex, and the sign of $N$ reflects the direction of the magnetic field.

\section{Solution for a particle bound to the vortex}\label{App:FreeParticle}

We now analyze the wavefunction under disk-shaped boundary conditions. The surrounding superconductor induces a finite gap in the system, allowing the central non-superconducting region to be treated as an antidot that hosts bound states. In our model, the energy scale of the normal-state dispersion in the antidot region, given by $\omega_\text{SC}({\bf k} \rightarrow \Gamma, \Delta = 0)$, satisfies $|\omega_\text{SC}| \ll |\Delta_0|$, where $\Delta_0$ denotes the superconducting gap.  %\rrc{I think this (if written as $|\omega_\text{SC}| \ll |\Delta_0|$) is the same as $|\mu|\ll |\Delta_0|$, that I wrote in a comment in section IIIC. This seems correct. I would still prefer the version with $\mu$, though. }
This condition ensures that 
the particle and hole branches in  Nambu space have small overlap and we can effectively describe the wave function by a single component. 
At the same time, the normal region acts as an effective potential well for quasiparticle confinement, supporting discrete in-gap states localized within the antidot.

In the presence of a magnetic field in the antidot, the momentum operator is given by ${\bf p} \rightarrow  -i\hbar  \nabla + |e|{\bf A}({\bf r})$,  and the Schr\"{o}dinger equation with the particle mass $m_b$ is given by
\begin{equation}
\bigg[\frac{1}{2m_b} \big(-i\hbar \nabla +|e|{\bf A}({\bf r})\bigr)^2 + V({\bf r}) \bigg]%\psi(r,\varphi) = E\psi(r,\varphi) 
\psi({\bf r}) = E_b\psi({\bf r})
\end{equation}
In the polar coordinate, ${\bf A}({\bf r}) = A_r \hat{\varphi}$, and the equation becomes
\begin{equation}
\begin{split}
&\Bigg[-\frac{\hbar^2}{2m_b}\frac{1}{r^2}\bigg\{ r\frac{\partial }{\partial r} \Big( r\frac{\partial }{\partial r} \Big)+  \frac{\partial^2 }{\partial \varphi^2} \bigg\} +\frac{1}{2m_b}e^2 A_r^2\\
& -2i\hbar |e| A_r  \frac{1}{2m_b}\frac{1}{r}\frac{\partial}{\partial \varphi}  \Bigg]  \psi (r, \varphi) = E_b \psi (r, \varphi). 
\end{split}
\end{equation}

%\red{JV: Using $R$ for the wave function is confusing since $R$ denotes the radius everywhere else (see B4 right below). }
We assume solution of the form $\psi(r, \varphi) = \mathcal{R}(r) \mathrm{e}^{i l\varphi}$ and use the form of vector potential inside the vortex. Substituting the solution in the Schr\"odinger equation 
for the radial component is given by 
\begin{equation}
 r^2\frac{\partial^2 \mathcal{R}(r)}{\partial r^2} + r \frac{\partial \mathcal{R}(r)}{\partial r} + \Bigg\{ r^2k^2 -\bigg(l+\frac{|e|A_r r}{\hbar}\bigg)^2 \Bigg\} \mathcal{R}(r) = 0 
\end{equation}
Here $k = \frac{1}{\hbar} \sqrt{2m_bE_b}$, and $l$ is the orbital angual momentum of the bound particle, as the system has the rotational symmetry. 
%and $m_b$ is the constant from the angular part and substitute $\nu = l+\tfrac{|e|A_r r}{\hbar}$. 

The vector potential in polar coordinate for magnetic field along the $z$ are given by, 
\begin{equation}
 A_r  =  \begin{cases}
&\frac{r}{2\pi R^2} \Phi  %~\hat{\varphi}  
\qquad ~\text{for} ~r\le R\\
&\frac{1}{2\pi r}\Phi %~\hat{\varphi} 
\qquad\quad ~\text{for} ~r> R 
 \end{cases}
\end{equation}

As in the main text, we consider $\Phi = N \Phi_0$. Using the form of the vector potential, the solution for the radial part is given by, 
\begin{equation}\label{eq:nu}
l'  = \begin{cases}
l + \frac{|e|}{2\pi\hbar} \bigl(\frac{r}{R}\bigr)^2 \Phi  \qquad \text{for} ~ r\le R\\
l+ \frac{|e|}{2\pi \hbar} \Phi   \qquad \qquad \text{for} ~r > R
\end{cases}
\end{equation}

In the absence of a magnetic field ($N = 0,~i.e.~\Phi =0$), the solutions to the radial Schrödinger equation correspond to the roots of the Bessel function, $\mathcal{J}_l(kr)$, since $l'=l$ in this case.   To satisfy the boundary condition that the wavefunction vanishes at the vortex edge ($r = R$), the allowed wavevectors are quantized as $k R = x_{n l }$, where $\mathrm{x}_{n l}$ is the $n^\text{th}$ zero of $\mathcal{J}_l(kr)$.  
The corresponding bound-state energies are given by %\jv{Eq.~(\ref{eq:Ebnl})} \red{JV: The below equation and 2 paragraphs seem to just repeat what was in the main text, so I think they could be removed from here: }
\begin{equation}\label{eq:FreePenergies}
    E_b^{n l} = \frac{\hbar^2}{2m_b} \frac{\mathrm{x}_{n l}^2}{R^2}.
\end{equation}

For zero flux, the angular momentum quantum number $l = 0$ yields a single energy level, while states with $l > 0$ exhibit a twofold degeneracy due to angular symmetry.

Figure~\ref{fig:Vortex_swave}~(b) compares these analytic energy levels from Eq.~\eqref{eq:FreePenergies} with numerical results obtained from a lattice model of the SC layer hosting an antidot with radius $R = 10$.  
To match the continuum results with the lattice data, we use $\hbar^2/(2m R^2) = 0.15\Delta_0$ as a fitting parameter.

This bound-state picture accurately captures the in-gap states of the lattice model, with the exception of an additional twofold spin degeneracy that is explicitly included in the lattice formulation but is absent in the scalar continuum model considered here.

As a result, all the lattice model eigenenergies show at least twofold degeneracy due to spin.  For $l \ne 0$, a further twofold degeneracy appears, consistent with the Bessel-function solutions in the zero-flux ($\Phi = 0$) limit. \\

For the finite-flux case ($N= 1$), the analysis becomes more subtle due to the modified differential equation governing the system, as discussed in Eq.~\ref{eq:nu}.  
Outside the vortex core, the solutions remain Bessel functions but with a shifted angular momentum index, reflecting the presence of magnetic flux.  
Inside the vortex, however, the solutions deviate from standard Bessel functions. Nonetheless, inside the antidot $r<R$, and limiting the $l<l'<l+1 $, with larger contribution from the antidot boundary. For simplicity assuming the mean radial distance $\langle r \rangle \rightarrow R$, %\rrc{The units don't match. And what 'radius' does this sentence refer to?}, 
the solutions in that region can therefore be approximately described by Bessel functions. %\rrc{This is a bit confusing. Since in this particle-in-a-disk model we demand $\mathcal{R}(R)=0$, the vector potential outside of the antidot should play no role. Inside the antidot, if the wavefunction lies in the region where $(r/R)^2$ is approximately constant, we can approximate the solution using constant $l'$. It seems though that it is when $\langle r \rangle$ of the eigenstate is small, we can approximate $l\approx l'$ and the solutions remain similar to the no-flux Bessel functions. }

Importantly, the degeneracy associated with the angular momentum quantum number $l$ is lifted for $l \neq 0$, consistent with the breaking of time-reversal symmetry by the magnetic field threading the vortex.  

Figure~\ref{fig:Vortex_swave}~(c) compares these analytical predictions with the eigenenergies obtained from the lattice simulation of the SC layer containing an antidot with radius $R = 10$ and a magnetic flux quantum $\Phi$ %\red{JV: What is $a$? The lattice constant hasn't been introduced in the introduction at least.}.  
The results clearly demonstrate the lifting of degeneracy for finite $l$ values, as expected from the theoretical considerations above.

%Further one can estimate the mass of the free-particles trapped in the vortex. Using, $a = 0.5$ nm, zeros of $\mathcal{J}_0(r)$ as 2.54, $\hbar = 1.05\times 10^{-34}$, lowest $E = 10^{-3}$ we estimate $m = 0.8 m_e $, where $m_e$ is the mass of the free-electron. This contrast with the low energy flat band considered for the SC layer, suggesting $m\rightarrow \infty$.
%Also, from the form of the equation, it is clear that the finite $l$ will split for a finite magnetic flux.

\bibliography{references}% Produces the bibliography via BibTeX.

\end{document}